\documentclass[a4paper,11pt]{article}
\pdfoutput=1 

\usepackage{jcappub} 

\usepackage[T1]{fontenc} 
\usepackage{epsf}
\usepackage{epstopdf}
\usepackage[usenames]{color}

\newcommand{\MeV}{\mathrm{MeV}}

\newcommand{\comment}[1]{}

\title{\boldmath Cosmic QCD phase transition: from quark to strangeon and nucleon?}

\author[a,b,c]{Xuhao Wu}
\author[b,c]{Weibo He}
\author[b,c]{Yudong Luo}
\author[d,e]{Guo-Yun Shao}
\author[b,c]{and Renxin Xu}

\affiliation[a]{State Key Laboratory of Metastable Materials Science and Technology $\&$ Key Laboratory for Microstructural Material Physics of Hebei Province, School of Science, Yanshan University, Qinhuangdao, 066004, China}
\affiliation[b]{School of Physics, Peking University, Beijing 100871, China}
\affiliation[c]{Kavli Institute for Astronomy and Astrophysics, Peking University, Beijing 100871, China}
\affiliation[d]{School of Science, Xi'an Jiaotong University, Xi'an, 710049, China}
\affiliation[e]{MOE Key Laboratory for Nonequilibrium Synthesis and Modulation of Condensed Matter,Xi'an Jiaotong University, Xi'an, 710049, China}
\emailAdd{wuhaobird@gmail.com}
\emailAdd{webb$\_$he@stu.pku.edu.cn}
\emailAdd{yudong.luo@pku.edu.cn}
\emailAdd{gyshao@mail.xjtu.edu.cn}
\emailAdd{r.x.xu@pku.edu.cn}

\abstract{
A crossover QCD phase transition in the early Universe, involving a scenario of forming stable strangeon nuggets is studied. The 2$+$1 Polyakov-Nambu-Jona-Lasinio model is applied to calculate the thermodynamics of the quark phase, and the relativistic mean-field model describes the hadronic one. The transition from quarks to hadrons occurred at a cosmic temperature of $T\sim170$ MeV, and these two phases are connected in a three-window model. 
It is proposed that, due to the non-perturbative coupling, strange quark clusters with net strangeness (i.e., strangeons) could form during the transition process, and these clusters can further grow to strangeon nuggets.
A distribution function of the nugget baryon number is introduced to describe the nuggets' number density.
All the strangeon nuggets with baryon number beyond $A_c$ are supposed to be stable, where the critical number, $A_c$, is determined by both the weak and strong interactions. A non-relativistic equation of state is applied to calculate the thermodynamics of stable strangeon nuggets, resulting in negligible thermodynamical contributions (pressure, entropy, etc.). The resultant mass density of the strangeon nuggets survival from 
the early Universe is comparable to that of dark matter, which indicates a possible explanation of the cold dark matter without introducing any exotic particles beyond the standard model.
}


\arxivnumber{2212.03466}

\begin{document}
\maketitle
\flushbottom

\section{Introduction}
\label{sec:1}

Cosmic phase transitions~\cite{Linde:1978px,Kibble:1980mv,Mazumdar:2018dfl,Hindmarsh:2020hop} are natural consequences of (hot) Big Bang cosmology, which could be essential to understand various cosmological phenomena such as primordial magnetic field~\cite{Durrer:2013pga}, baryonic asymmetry~\cite{Kuzmin:1985mm,Farrar:1993sp}, and even gravitational wave background~\cite{Kosowsky:1992rz,Schettler:2010dp}.
At least two types of transitions occurred in the early Universe: the quantum chromodynamical (QCD) phase transition~\cite{Boeckel:2010bey} and the electroweak (EW) one~\cite{Cline:2008hr}.
The latter happens when the cosmic temperature drops below $T \sim 10^2$ GeV at which the electroweak symmetry is broken, allowing the Standard Model particles to acquire gauge invariant masses~\cite{Higgs:1964pj,Weinberg:1974hy}, while for the former,  the exact dynamics is still unclear since the perturbative theory breaks down during the QCD phase transition.
Stable quark nuggets with large baryon number may survive if the QCD transition is of first-order~\cite{Witten1984}, but the transition could be a rapid and
smooth crossover~\cite{HotQCD:2014kol,Schmidt:2017bjt,Aarts2023} in lattice QCD and many effective QCD models.
Nevertheless, one of the most interesting questions relevant to cosmic QCD separation is: could the real nature of dark matter be the strong matter nuggets surviving from the early Universe?
This issue has been fully focused on since the work by Witten~\cite{Witten1984} in 1984, and 
this kind of dark matter candidate could even help in explaining the puzzling supermassive black holes at high redshifts~\cite{Peebles2022} by invoking a quick formation of seed black hole~\cite{Lai:2010JCAP}.
It is, therefore, the focus of this work, to investigate further such a transition and the accompanying dark matter production of strong nuggets in a scenario of crossover QCD phase transition.

\begin{figure}[h]
\centering
\includegraphics[ width=16 cm,clip]{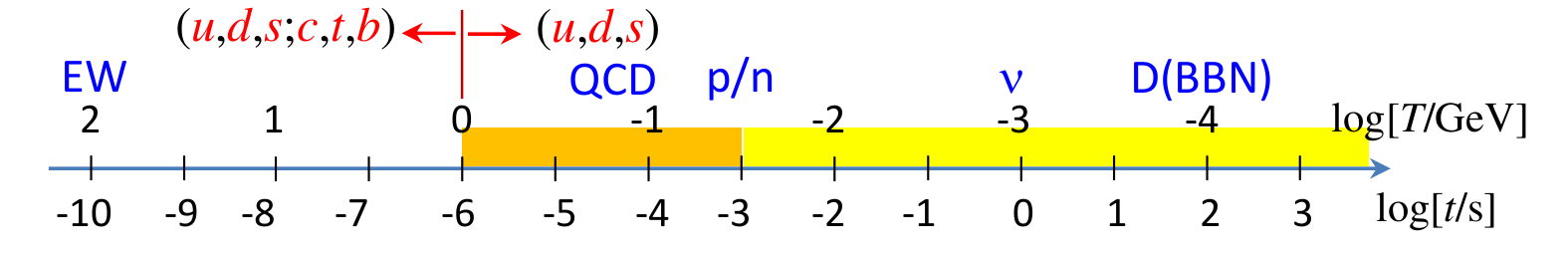}%
\caption{Cosmic evolution from the electroweak (EW) epoch to the big bang nucleosynthesis (BBN). Free nucleons (protons and neutrons, $p$/$n$) may appear after QCD phase transition at $T\sim100$ MeV, setting the initial condition of BBN. Neutrinos ($\nu$) are decoupled at $T\sim1$ MeV, while the deuterium (D) forms at $T\sim 0.1$ MeV.}
\label{fig:1time}
\end{figure}

After the EW phase transition, the Universe was fulling with dense hot quark-gluon plasma (QGP) with six flavors of quarks until the temperature dropped to a few giga-electronvolts. Then, the heavy flavors of quarks, i.e., charm ($c$), top ($t$), and bottom ($b$) quarks (bare masses $>$ 1 GeV), started to decay to light ones: up ($u$), down ($d$) and strange ($s$) quarks (masses $\lesssim$ 0.1 GeV) under the chemical equilibrium. After that, the Universe comes to the QCD phase transition. During this epoch, the initial strongly-interacting quark matter phase at $T\geq$ 100 MeV will end up with the hadronic matter (e.g., nucleons) phase at $T\sim$ 10 MeV, providing the initial conditions of the big bang nucleosynthesis (BBN). The cosmic timeline of this epoch is shown in Fig.~\ref{fig:1time}~\cite{Kolb1990,Schwarz2003}.

For a first-order cosmological QCD phase transition, the low-temperature bubbles appear after the cosmic temperature drops below the critical temperature $T_c \sim (100-200)$ MeV \cite{Witten1984}. They expel the heat to the surrounding high-temperature phase, slowly expand and collide. As the Universe expands and the temperature decreases, at some stage, the dense and high-temperature bubbles become isolated. They would lose energy due to surface evaporation and neutrino emission. The latter carries only the leptons out, leaving the baryons inside the bubbles. Therefore eventually, the baryon excess inside bubbles would become stable. 

However, it is still a matter of debate whether the cosmic QCD phase transition is of first-order. A smooth crossover phase transition could be reasonable, starting from the QGP phase with free quarks to hadrons, as discussed in lattice QCD and many effective QCD models \cite{HotQCD:2014kol,Schmidt:2017bjt,Aoki:2009sc}. 
At a cosmic temperature above $100$ MeV, three flavors of quarks ($u$,$d$,$s$) exist simultaneously in the equilibrium state. 
These quarks could collide during the crossover QCD phase transition, and  nucleon-like quark clusters with strangeness (so-called ``strangeon'' \cite{Xu2019}) could then form via nucleation. Strangeons would continue to merge and smash to form nuggets.
These strangeon nuggets may evaporate particles, such as strangeon, $\Lambda$, and nucleon, decaying finally into neutrons and protons~\cite{Lai2021} at cosmic temperature above $\sim$ 10 $\MeV$~\cite{Alcock1985, Madsen1986}. With temperature decreases, evaporation will be suppressed so that the rest of the strangeon nuggets will become thermodynamically stable. Strangeon nuggets may then survive if they contain enough baryon numbers (it is worth noting that the readers should distinguish the strangeon nuggets from the strangelet proposed by Ref. \cite{Witten1984}, the latter is a group of $u,d$ and $s$ quarks, and it could be stable only if the baryon number inside the strangelets is about $\sim 10^{44}$ at the beginning~\cite{Sumiyoshi1991, Bhattacharjee1993}). Such stable nuggets are an analogy with the ordinary atomic nucleus: neutrons should decay into protons, but nuclei are stable due to interactions (both the strong and the weak) between protons and neutrons inside. As for the strangeon nuggets, the basic unit of the nuggets is strangeons, and the heavy nuggets could also be stable due to the weak and strong interactions \cite{Xu2003}. 

Many previous studies investigated the strangeon matter from the astrophysical perspective~\cite{Lai2023}, aiming to solve the problem raised by Lev Landau more than ninety years ago~\cite{Xu2023}. In fact, the strangeon stars could be considered as huge strangeon nuggets with stellar size. Some pulsar observations may indicate the existence of strangeon stars. For example, previous works \cite{Xu:1999bw,FAST:2019zow} suggested the bare strangeon stars could explain the sub-pulse drift signal, Refs.~\cite{Lai:2017xys, Wang:2020xsm} showed further the potential of strangeon stars to explain the observed glitch amplitude. 
On the other hand, several theoretical researches\cite{Zhou:2004ue,Peng:2007eq, Zhou:2014tba} studied the pulsar glitches mechanism in the strangeon star model. The global parameters of  non-rotating and rotating strangeon stars, as well as the oscillation modes, have also been investigated in Ref. \cite{Gao:2021uus} and Ref. \cite{Li:2022qql}.
In addition, as for the strange quark matter formed in the early Universe, a recent study discusses the possibility of destroying primordial $^7$Li  abundance via a 2 MeV photon emission line from color superconducting quark nuggets \cite{Ouyed:2023hqe}. Our previous work \cite{Lai:2010JCAP} investigated the possibility that strangeon nuggets formed during the first-order QCD phase transition could collapse to a stellar-mass black hole, then it kept growing by the gas accretion and became the supermassive black hole at redshift $z>6$.

This work considers the formation of strangeon nuggets during the crossover phase transition in the early Universe. The $s$-quark itself is not stable, decaying via $s \to e^- + u + \bar{\nu}_e$ with lifetime $\sim 10^{-9}$ s. However, a large strangeon nugget would be stable since a huge number of $s$-quarks cannot decay simultaneously into $u/d$-quarks via the weak interaction. %
In fact, there is a threshold baryon number $A_c$ for the baryon evaporation, and small strangeon nuggets with $A<A_c$ will be completely destroyed by the weak interaction or evaporation.
For a crossover phase transition, the strangeon nuggets are formed via collision and nucleation, so their baryon number may not be as large as produced during the first-order phase transition ($A_c\sim10^{44}$ for the first-order phase transition). 
The large nuggets would interact negligibly with normal baryonic matter via strong, weak, or electromagnetic interactions, i.e., they are a potential candidate for cold dark matter (CDM) \footnote{Small nuggets could collide with the nucleus during the primordial nucleosynthesis epoch. However, this fact does not rule out the possibility of a small value of $A_c$. Future studies on BBN network involving strangeon nuggets could provide more strict constraints on $A_c$ value.}.

In fact, there are several potential candidates of the CDM: axion, WIMPs (weakly interacting massive particles), primordial black holes, etc~\cite{Schwarz2003}. 
The axion and WIMPs are hypothetical elementary particles and are not in the standard model. Before the QCD transition, the required density fluctuation should be significant to produce primordial black holes in the very early Universe, as discussed in Refs.~\cite{Nadezhin1978, Carr1974, Schwarz2003}. Many of the proposals invoking physics beyond the standard model have been excluded, and the remaining allowed regions of parameter space are narrowing~\cite{Farrar2003, Farrar2022}. One of the motivations of this study is that the stable strangeon nuggets is a potential candidate of dark matter in the regime of ``old'' physics without invoking any exotic variety of particles. Actually, it has a long history~\cite{Witten1984, Zhitnitsky2003, Alonso-Alvarez2021, Flambaum2021, Farrar2022} to connect strangeness with dark matter because of a meager charge-to-mass ratio if the symmetry of light-quark flavors ({\em uds}) is restored.
%

This article is organized as follows.
In Section~\ref{sec:2}, we briefly introduce the Polyakov-Nambu-Jona-Lasinio (PNJL) model for quark matter in the QGP phase and the relativistic mean field (RMF) model for the hadron phase. In Section~\ref{sec:3},  we discuss the formation of the strangeon nuggets and the crossover QCD phase transition.
In Section \ref{sec:results}, we show
the numerical results of the QCD phase transition
with discussions. Section \ref{sec:summary} is devoted to a summary.


\section{Equations of state of Quark Phase and Hadron phase}
\label{sec:2}

\begin{figure}[b]
\centering
\includegraphics[ width=8 cm,clip]{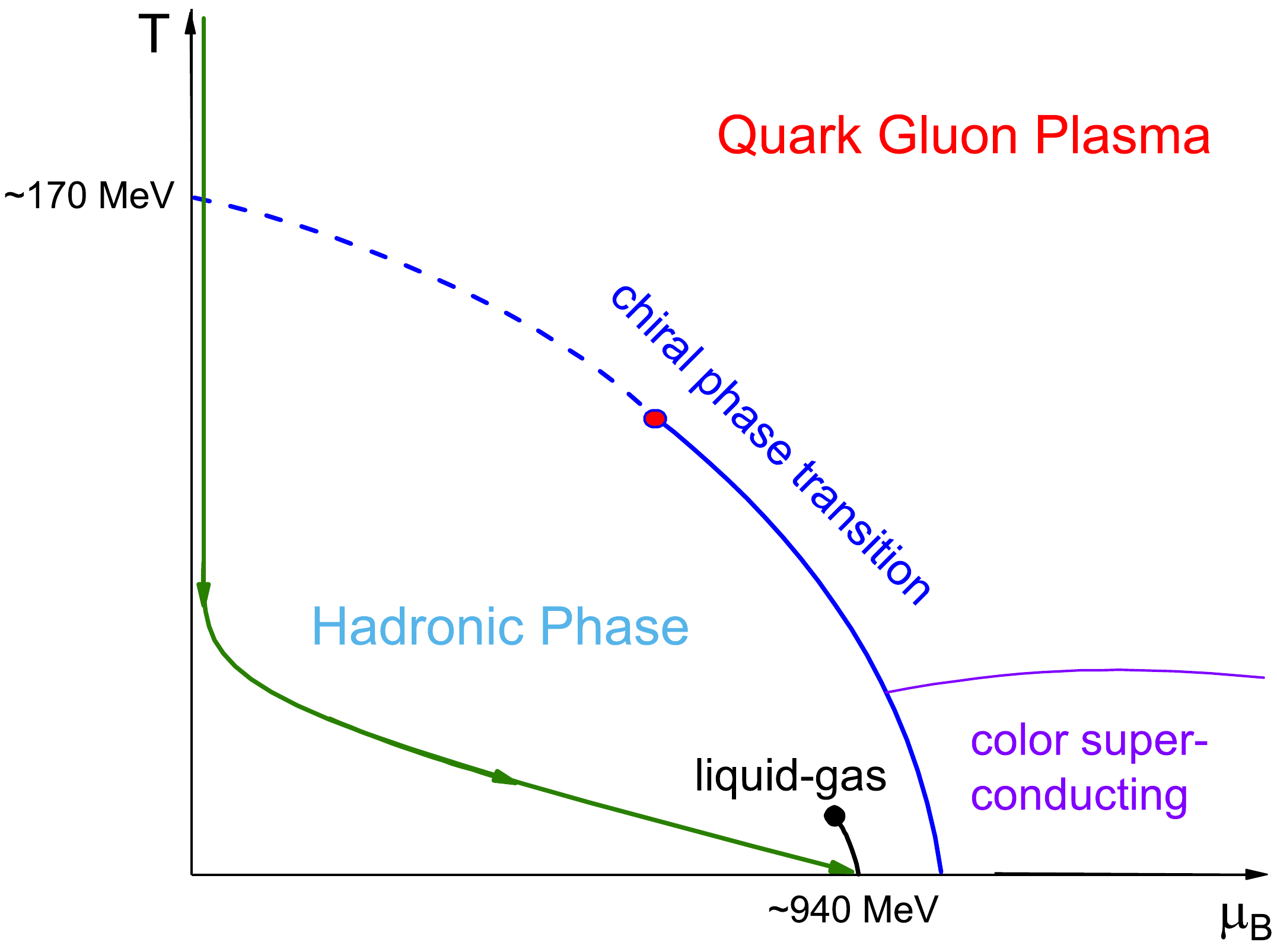}%
\caption{ Sketch of the QCD phase diagram in terms of temperature $T$ and baryon chemical potential $\mu_B$. The green curve with arrows illustrates the thermal trajectory of the universe in the QCD epoch.}
\label{fig:qcdiag}
\end{figure}

During the QCD phase transition, the statistical equilibrium and the charge neutrality are satisfied. A conjectured QCD phase diagram in the T-$\mu_B$ plane is presented in Fig. \ref{fig:qcdiag} (quantitative discussions of QCD phase diagram could be found in Ref. \cite{Guenther:2020jwe,Bzdak:2019pkr,Fischer:2018sdj}). The first-order nuclear liquid-gas phase transition (solid black line) occurs at a low temperature, and $\mu_B$ is approximately equal to nucleon mass. The hadron-quark phase transition is a crossover (blue dashed line) at high temperature and is first-order (solid blue line) at low temperature. They are connected by the critical end point (red dot). Although the exact curve that distinguishes the QGP phase and hadron gas phase is not yet determined, the present experiment and theory can still provide an overview of the diagram. In low temperatures and extremely large $\mu_B$, the color superconducting phase, various forms of quark Cooper pairing may appear. We consider a trajectory of the Universe in the QCD phase diagram to follow a crossover phase transition, which descends close to the vertical axis with almost zero chemical potential. Then at temperature $T\sim 100$ MeV starts to approach the nuclear matter region at low temperature $T\sim 1$ MeV.

In general, the equation of state (EOS) of the QGP phase and the Hadron-Strangeon nuggets (HS) phase under the finite temperature require two of the thermodynamical quantities: number density, temperature, entropy density, and lepton fraction $Y_l$. In this work, the thermodynamical quantities are set as functions of $T$ and $s/n_b$. The Universe is isentropic during the expansion, for relativistic particles ~\cite{Lineweaver2008, Kolb1990, Zemansky1997} that the entropy of the Universe is dominated by the relativistic gas, the entropy density $s$ is proportional to the number of particles:
\begin{equation}
  s\equiv {S\over V} = {\varepsilon +p \over T} = {2\pi^2 \over 45} g_{\ast s} T^3,
\end{equation}
here, $g_{\ast s}$ is the summation of the degree of freedom for all the relativistic particles. Since most of the time, these particles share the same temperature, the entropy density is proportional to the photon number density: $s = 1.80g_{\ast s} n_\gamma$. The entropy per baryon $s/n_b$ is a conserved quantity with respect to the co-moving frame of reference since $n_b\propto T^3$, the value of $s/n_b$ can be derived from the baryon-to-photon ratio $\eta$ by the relation as $\eta = 1.8 g_{\ast s} (n_b/s)$. The current cosmic microwave background power spectrum analysis constrains the baryon-to-photon ratio $\eta$ as $ (6.16 \pm 0.02) \times 10^{-10} $, which is corresponding to the baryon density $\Omega_b h^2  = 0.0224 \pm 0.0001$ in the standard $\Lambda$CDM model \cite{Planck}. In this work, we use the value of $s/n_b$ refers to this $\eta$ value, and for comparison, we choose another extreme opposite case with $s/n_b = 100$.


\subsection{Quark Phase}
\label{sec:Qphase}
The three-flavor PNJL model is constructed by SU(3) Nambu-Jona-Lasinio (NJL) model
coupled to a temporal background gauge field, which represents Polyakov loop dynamics. 
The effective Lagrangian is written as (hereafter, we use the natural unit $c=\hbar=k_B =1$)
\begin{eqnarray}
{{\cal L}_{{\rm{PNJL}}}} &=& \bar q \left( {i{\gamma _\mu }{D^\mu } - {m^0}} \right)q  \nonumber\\
&&+ {G}\sum\limits_{a = 0}^8 {\left[ {{{\left( {\bar q{\lambda _a}q} \right)}^2} + {{\left( {\bar qi{\gamma _5}{\lambda _a}q} \right)}^2}} \right]} \nonumber\\
&&- K\left\{ {\det \left[ {\bar q\left( {1 + {\gamma _5}} \right)q} \right] + \det \left[ {\bar q\left( {1 - {\gamma _5}} \right)q} \right]} \right\}  \nonumber\\
&&- {\cal U}\left( {\bar \Phi  ,\Phi ,T} \right) ,
\end{eqnarray}%
where $q$ denotes a quark field with three flavors $\left( N_{f}=3\right) $
and three colors $\left( N_{c}=3\right) $. 
$m^{0}=$diag$\left(
m_{u}^{0},m_{d}^{0},m_{s}^{0}\right) $ is the current quark mass matrix, 
and we assume isospin symmetry $m_{u}^{0}=m_{d}^{0}\equiv m_{q}^{0}$.
We use the parameters in Ref.~\cite{Rehberg1996}, $m_{q}^{0}=5.5 ~\MeV$, $m_s^0=140.7$ MeV, $\Lambda=603.2$ MeV, $G\Lambda^2=1.835$, $K\Lambda^5=12.36$.
The quantity ${\cal U}\left( {\bar \Phi  ,\Phi ,T} \right)
$ is the effective potential in terms of $\bar \Phi$ and $\Phi$,
\begin{eqnarray}
\Phi  &=& \left( {T{r_c}L} \right)/{N_c}, \\
\bar \Phi   &=& \left( {T{r_c}{L^ + }} \right)/{N_c} ,
\end{eqnarray}%
which are the traced Polyakov loop and its conjugate.
A logarithmic formed Polyakov loop potential ${\cal U}\left(\Phi ,\bar \Phi  ,T \right)$ is applied~\cite{Ratti2006}
\begin{eqnarray}
\frac{{\cal U}\left(\Phi ,\bar \Phi  ,T \right)}{T^4} &=&  - \frac{{{b_2}\left( T \right)}}{2}\bar \Phi  \Phi - {b_4}\left( T \right)\ln \left[ 1 - 6\bar \Phi  \Phi \right. \nonumber\\
&& \left.+ 4\left( \bar{ \Phi}^3 + {\Phi ^3} \right) - 3\left( {\bar\Phi  \Phi } \right)^2\right]  ,
\label{eq:Poly2}
\end{eqnarray}%
with
\begin{eqnarray}
{b_2}\left( T \right) &=& {a_0} + {a_1}\left( {\frac{{{T_0}}}{T}} \right) + {a_2}{\left( {\frac{{{T_0}}}{T}} \right)^2}   ,\nonumber\\
{b_4}\left( T \right) &=& {b_4}{\left( {\frac{{{T_0}}}{T}} \right)^3}  .
\label{eq:Poly4}
\end{eqnarray}%
The parameters are given in Table.~\ref{tab:rrw06}. In the table, $T_0$ is characterized by the jump of $\Phi$ from the vanishing to a finite value.
\begin{table}[htp]
  \caption{Dimensionless parameters of the potentials and $T_0$ given in
Eqs.~(\ref{eq:Poly2}), (\ref{eq:Poly4})~\cite{Ratti2006}. }
  \begin{center}
    \setlength{\tabcolsep}{2.6mm}{
      \begin{tabular}{lcccccccccccc}
        \hline\hline
        $a_0$   &$a_1$  &$a_2$  &$a_3$  &$b_3$ &$b_4$   & $T_0$  \\
        \hline
        6.75   &-1.95  &2.625  &-7.44  &0.75    &7.5  &270 MeV    \\
        \hline\hline
    \end{tabular}}
    \label{tab:rrw06}
  \end{center}
\end{table}

The constituent quark mass obeys the gap equation
\begin{eqnarray}
m_i = {m_{i}^0} - 4G{\phi_i}+2K{\phi_j}{\phi_k} ,
\end{eqnarray}%
where $\phi_i$ is the quark condensate. $\Phi$ and $\bar{\Phi}$ are given by
\begin{eqnarray}
\frac{\partial \Omega}{\partial \phi_{i}}&=&0, \\
\frac{\partial \Omega}{\partial \Phi}&=&0, \frac{\partial \Omega}{\partial \bar{\Phi}}=0.
\end{eqnarray}
$\Omega$ is the grand canonical potential
\begin{eqnarray}
\Omega&=& U(\bar{\Phi}, \Phi, T)+2 G\left(\phi_{u}^{2}+\phi_{d}^{2}+\phi_{s}^{2}\right)-4 K \phi_{u} \phi_{d} \phi_{s} 
\nonumber\\
&&-2 N_c\int_{\Lambda} \frac{\mathrm{d}^{3} k}{(2 \pi)^{3}}\left(E_{u}+E_{d}+E_{s}\right) 
\nonumber\\
&&-2\sum_{i=u, d, s} \int \frac{\mathrm{d}^{3} k}{(2 \pi)^{3}}\frac{k^2}{E_i^2}[F^+(E_i-\mu_i,T,\Phi,\bar\Phi)
\nonumber\\
&&\left.+F^-(E_i+\mu_i,T,\Phi,\bar\Phi)\right],
\end{eqnarray}
with
\begin{align}
 {F^ + }\left( {x,T,\Phi ,\bar \Phi  } \right) = \frac{{\Phi {e^{ - x /T}} + 2\bar \Phi  {e^{ - 2{x/T}}} + {e^{ - 3x/T}}}}{{1 + 3\Phi {e^{ -x/T}} + 3\bar \Phi  {e^{ - 2x/T}} + {e^{ - 3x/T}}}}, \\
 {F^ - }\left( {x,T,\Phi ,\bar \Phi  } \right) = \frac{{\bar \Phi  {e^{ - x/T}} + 2\Phi {e^{ - 2x/T}} + {e^{ - 3x/T}}}}{{1 + 3\bar \Phi  {e^{ - x/T}} + 3\Phi {e^{ - 2x/T}} + {e^{ - 3x/T}}}},
\end{align}
being the generalized Fermi-Dirac distribution and ${E_i} = \sqrt{k^2 + m^2}$. 
Through grand canonical potential, all the thermodynamic quantities can be obtained.
These are the expressions of the pressure $P$
\begin{eqnarray}
\label{eq:pnjlp}
 P_{{\rm{PNJL}}}  &=&   -{\cal U}\left( {\bar \Phi  ,\Phi ,T} \right) 
 \nonumber\\
 &&- 2G\left({\phi_u^2}+{\phi_d^2}-{\phi_s^2}\right)
 +4K{\phi_u}{\phi_d}{\phi_s}  \nonumber\\
   &&+ 2{N_c}{N_f}\frac{1}{3}\int_0^\infty  {\frac{{{k^2}dk}}{{2{\pi ^2}}}} \frac{{{k^2}}}{{{E_i}}}\left( {{F^ + } + {F^ - }} \right)
  \nonumber\\
&& + 2{N_c}{N_f}\int_0^\Lambda  \frac{k^2dk}{2\pi ^2} {E_i} + \varepsilon _{\rm{vac}},
\end{eqnarray}
the quark density
\begin{eqnarray}
\label{eq:pnjln}
n &=& {\rho _v} = 2{N_c}{N_f}\int_0^\infty  {\frac{{{k^2}dk}}{{2{\pi ^2}}}} \left( {{F^ + } - {F^ - }} \right)
,
\end{eqnarray}%
the energy density
\begin{eqnarray}
\label{eq:pnjle}
 \varepsilon _{\rm{PNJL}} &=& {\cal U}\left( {\bar \Phi  ,\Phi ,T} \right) - T\frac{{\partial {\cal U}}}{{\partial T}} 
 \nonumber\\
 &&+ 2G\left({\phi_u^2}+{\phi_d^2}-{\phi_s^2}\right)
 -4K{\phi_u}{\phi_d}{\phi_s} 
 \nonumber\\
   &&+ 2{N_c}{N_f}\int_0^\infty  {\frac{{{k^2}dk}}{{2{\pi ^2}}}} {E_i}\left( {{F^ + } + {F^ - }} \right)
  \nonumber\\
 &&- 2{N_c}{N_f}\int_0^\Lambda  \frac{k^2dk}{2\pi ^2} {E_i} - \varepsilon _{\rm{vac}},
  \end{eqnarray}%
and the entropy density
\begin{eqnarray}
\label{eq:pnjls}
 s_{\rm{PNJL}} &=&  - \frac{{\partial {\cal U}}}{{\partial T}} + 2{N_c}{N_f}\frac{1}{3}\frac{1}{T}\int_0^\infty  {\frac{{{k^2}dk}}{{2{\pi ^2}}}} \frac{{{k^2}}}{{{E_i}}}\left( {{F^ + } + {F^ - }} \right) \nonumber\\
 && + 2{N_c}{N_f}\frac{1}{T}\int_0^\infty  {\frac{{{k^2}dk}}{{2{\pi ^2}}}} {E_i}\left( {{F^ + } + {F^ - }} \right)  \nonumber\\
 &&- 2{N_c}{N_f}\frac{\mu }{T}\int_0^\infty  {\frac{{{k^2}dk}}{{2{\pi ^2}}}} \left( {{F^ + } - {F^ - }} \right) .
\end{eqnarray}%

Inside the homogeneous matter, particles (and antiparticles) occupy single-particle states with the Fermi-Dirac distribution. The particle and antiparticle occupation probability is given by
\begin{eqnarray}
f_{i}^{k} &=&\left[ 1+e^{ (E_{k}^{i}-\mu_i )/T}\right]^{-1}, \\
f_{\bar{i}}^{k} &=&\left[ 1+e^{ (E_{k}^{i}+\mu_i )/T}\right]^{-1},
\end{eqnarray}%
 respectively.
 The leptons are considered as free Fermi gas, which has
 \begin{eqnarray}
P_{l} &=&\frac{1}{3}\sum\limits_{l = e,\mu } {\frac{1}{{{\pi ^2}}}\int  {\frac{{{k^4}}}{{{{\left( {{k^2} + m_l^2} \right)}^{1/2}}}}\left( {f_l^k + f_{\bar l}^k} \right)dk} } , \\
\epsilon_{l} &=&  \sum\limits_{l = e,\mu } {\frac{1}{{{\pi ^2}}}\int  {{{\left( {{k^2} + m_l^2} \right)}^{1/2}}{k^2}\left( {f_l^k + f_{\bar l}^k} \right)dk} } , \\
s_{l} &=& \sum\limits_{l =e,\mu} {\frac{1}{{{\pi ^2}}}\int {dk\left[ { - f_l^k\ln } \right.f_l^k - \left( {1 - f_l^k} \right)\ln \left( {1 - f_l^k} \right)} }  \nonumber\\
&& \left. { - f_{\bar l }^k\ln f_{\bar l }^k - \left( {1 - f_{\bar l }^k} \right)\ln \left( {1 - f_{\bar l }^k} \right)} \right].
\end{eqnarray}%
Adding the contribution of leptons, we have
\begin{eqnarray}
 P_{{\rm{Q}}}   &=&   P_{{\rm{PNJL}}} +P_{l}, \\
 \epsilon_{{\rm{Q}}}   &=&   \epsilon_{{\rm{PNJL}}} +\epsilon_{l}, \\
 s_{{\rm{Q}}}   &=&   s_{{\rm{PNJL}}} +s_{l}.
\end{eqnarray}%

\subsection{Hadron Phase}
\label{sec:Hphase}
To describe the hadronic matter, we use the RMF theory, which conforms to experimental data and saturation properties well under low densities.
The Lagrangian reads
\begin{eqnarray}
\label{eq:RMF}
\mathcal{L}_{\rm{RMF}} & = & \sum_{i=p,n}\bar{\psi}_i
\bigg \{i\gamma_{\mu}\partial^{\mu}-\left(M+g_{\sigma}\sigma\right)
\notag\\
&&-\gamma_{\mu} \left[g_{\omega}\omega^{\mu} +\frac{g_{\rho}}{2}\tau_a\rho^{a\mu}
\right]\bigg \}\psi_i  \notag \\
&& +\frac{1}{2}\partial_{\mu}\sigma\partial^{\mu}\sigma -\frac{1}{2}%
m^2_{\sigma}\sigma^2-\frac{1}{3}g_{2}\sigma^{3} -\frac{1}{4}g_{3}\sigma^{4}
\notag \\
&& -\frac{1}{4}W_{\mu\nu}W^{\mu\nu} +\frac{1}{2}m^2_{\omega}\omega_{\mu}%
\omega^{\mu}   \notag
\\
&& -\frac{1}{4}R^a_{\mu\nu}R^{a\mu\nu} +\frac{1}{2}m^2_{\rho}\rho^a_{\mu}%
\rho^{a\mu}
\notag\\
&& +\sum_{l=e,\mu}\bar{\psi}_{l}
  \left( i\gamma_{\mu }\partial^{\mu }-m_{l}\right)\psi_l,
\end{eqnarray}
where
\[\begin{array}{l}
 {W^{\mu \nu }} = {\partial ^\mu }{\omega ^\nu } - {\partial ^\nu }{\omega ^\mu } ,\\
 {R^{\alpha \mu \nu }} = {\partial ^\mu }{\omega ^{\alpha \nu }} - {\partial ^\nu }{\omega ^{\alpha \mu }} + {g_\rho }{\varepsilon ^{\alpha \beta \gamma }}{\rho ^{\beta \mu }}{\rho ^{\gamma \nu }} .\\
 \end{array}\]
In this RMF model, the interactions between hadrons are represented by exchanging mesons: $\sigma$ meson reflects the mid-range attraction; the $\omega$ meson represents the short-range repulsion; the $\rho$ meson represents the isospin difference between neutron and proton. The mass of the mesons and the coupling constants are given in Table~\ref{tab:rmf}. $W^{\mu \nu }$ and $R^{\alpha \mu \nu }$ are the antisymmetric field tensors.
\begin{table*}[htp]
  \caption{Parameters in the GM1 model~\cite{Glendenning1991}.
    The masses are given in MeV. }
  \begin{center}
    \setlength{\tabcolsep}{1.6mm}{
      \begin{tabular}{lcccccccccccc}
        \hline\hline
        Model   &$M$  &$m_{\sigma}$  &$m_\omega$  &$m_\rho$  &$g_\sigma$  &$g_\omega$
        &$g_\rho$ &$g_{2}$ (fm$^{-1}$) &$g_{3}$   \\
        \hline
        GM1     &938.000  &550.000  &783.000  &770.000  &9.5705  &10.6096  &8.1954
        &$-$12.2799   &-8.9767     \\
        \hline\hline
    \end{tabular}}
    \label{tab:rmf}
  \end{center}
\end{table*}

Hadronic matter satisfies the statistical equilibrium and charge neutrality condition, i.e.,
\begin{eqnarray}
 {\mu _n} &=& {\mu _p} + {\mu _e},\\
 {\mu _\mu } &=& {\mu _e},\\
\label {chemical potential}
 {n_p} &=& {n_e} + {n_\mu }
\label {electric neutrality}.
\end{eqnarray}

The vector density is
\begin{eqnarray}
 {n_i} &=& {\rho _v} = \left\langle {{{\bar \psi }_b}{\gamma _0}{\psi _b}} \right\rangle    \nonumber\\
 &=& \frac{1}{{{\pi ^2}}}\int {{k^2}\left( {f_i^k - f_{\bar i }^k} \right)dk} ,
 \end{eqnarray}
and the scalar density is given by
\begin{equation}
{\rho _s} = \left\langle {{{\bar \psi }_b}{\psi _b}} \right\rangle   = \frac{1}{{{\pi ^2}}}\int {\frac{{m_N^*}}{{{E^*}}}{k^2}\left( {f_i^k + f_{\bar i }^k} \right)dk} .
\end{equation}
The energy density, the pressure, and the entropy density of the hadron phase read
\begin{eqnarray}
 \varepsilon_{{\rm{H}}}  &=& {\kern 1pt} {\kern 1pt} \frac{1}{2}m_\sigma ^2{\sigma ^2} + \frac{1}{3}{g_2}{\sigma ^3} + \frac{1}{4}{g_3}{\sigma ^4} \nonumber\\
 &&+ \frac{1}{2}m_\omega ^2{\omega ^2} + \frac{3}{4}{c_3}{\omega ^4} + \frac{1}{2}m_\rho ^2{\rho ^2} \nonumber\\
  &&+ \sum\limits_{i = n,p} {\frac{1}{{{\pi ^2}}}\int  {{{\left( {{k^2} + m{{_N^ * }^2}} \right)}^{1/2}}{k^2}\left( {f_i^k + f_{\bar i}^k} \right)dk} } +\epsilon_{l}, \nonumber\\
\label{eq:eoshe}    \\
 P_{{\rm{H}}} &=& {\kern 1pt} {\kern 1pt}  - \frac{1}{2}m_\sigma ^2{\sigma ^2} - \frac{1}{3}{g_2}{\sigma ^3} - \frac{1}{4}{g_3}{\sigma ^4} \nonumber\\
 &&+ \frac{1}{2}m_\omega ^2{\omega ^2} + \frac{1}{4}{c_3}{\omega ^4} + \frac{1}{2}m_\rho ^2{\rho ^2} \nonumber\\
  &&+ \frac{1}{3}\sum\limits_{i = n,p} {\frac{1}{{{\pi ^2}}}\int  {\frac{{{k^4}}}{{{{\left( {{k^2} + m{{_N^ * }^2}} \right)}^{1/2}}}}\left( {f_i^k + f_{\bar i}^k} \right)dk} }+P_{l} ,  \nonumber\\
\
  \\
   s_{{\rm{H}}} &=& \sum\limits_{i = n,p} {\frac{1}{{{\pi ^2}}}\int {dk\left[ { - f_i^k\ln } \right.f_i^k - \left( {1 - f_i^k} \right)\ln \left( {1 - f_i^k} \right)} }  \nonumber\\
&& \left. { - f_{\bar i }^k\ln f_{\bar i }^k - \left( {1 - f_{\bar i }^k} \right)\ln \left( {1 - f_{\bar i }^k} \right)} \right] +s_{l} ,
\label{eq:eoshs}
\end{eqnarray}
separately.

\section{The formation of strangeon nuggets and crossover phase transition}
\label{sec:3}

We propose that the QCD phase transition is a continuous crossover phase transition. During the transition, quarks collide with each other and form nuggets. The small nuggets will decay quickly to baryons, while the large quark nuggets could survive after temperature decreases under $T\sim 100\ \rm MeV$. The nature of a strongly-interacting system, either nucleon-constituted nucleus or strangeon-constituted nuggets, is determined by the fundamental strong and the weak interactions.
From an astrophysical point of view, it is conjectured that bulk strangeon nuggets could be more stable even than the nucleus $^{56}$Fe~\cite{Lai2017}. Therefore, strangeon nuggets are long-lived if their baryon number $A$ is larger than a critical number $A_{\rm c}$, and they are massive enough to be considered as classical particles. In contrast, the nuggets with baryon number $A<A_{\rm c}$ will decay to neutrons quickly and establish a nuclear statistical equilibrium with the proton since the temperature is still higher than the weak interaction decoupling temperature $T_{dec} \sim 1\ \rm MeV$. The value of $A_{\rm c}$ is determined by the interaction scale. For example, suppose we only consider the weak decay of the $s$ quark may cause the instability of the strangeon droplets. In that case, the critical scale is given by the electron Compton wavelength, $D_{\rm c} \sim \lambda_{\rm c} = 2\pi/(m_{\rm e}c)=2.4\times 10^3$ fm, so we have  $A_c \sim 10^9$.  Alternatively, one has $A_{\rm c}\simeq 300$ if only the strong interaction is considered~\cite{Wang2018}. So in this work, we use the suggested value of $A_{\rm c}\simeq (10^3\sim 10^9)$. The realistic calculation needs to consider the hybrid existence of the quark and HS phase and integrate the transition process with the strangeon nugget fraction evolution. For simplicity, we use a distribution function to describe the co-existence state of hadrons and strangeon nuggets after the phase transition. The timescale of the QCD phase transition is $\sim 10^{-6}$ s. It is much longer than the relaxation timescale for the Universe to achieve the thermal equilibrium, so we define a crossover region near the critical temperature $T_c$ and use a smooth interpolation of the Helmholtz free energy per baryon to describe the continuous phase transition.

\subsection{Formation of the Strangeon Nuggets}
It is more difficult to form a nugget with a larger baryon number $A$ via collisions, so larger $A$ should have a smaller number density. Similar to the nuclei, the baryon number of a nugget is proportional to its volume: $A= (D/D_0)^3$, where $D$ is the diameter of the nuggets. In this work, an exponential distribution as a function of the size of the nuggets $D$ is assumed: 
\begin{eqnarray}
\label{eq:sndis}
n\left(D\right)=n_0 e^{-D/R_c} = n_0 e^{-{2D\over D_0A_c^{1/3}}},
\end{eqnarray}
where $R_c = D_c /2$ is the critical radius for the strangeon nuggets that could be stable. $n_0$ is the normalization factor since the total number density of the hadron should be:
\begin{equation}
  \int_0^{D_c} n_0 e^{-D/R_c} dD= 2\Big({mT\over 2\pi}\Big)^{3/2}\exp{\Big[-(m-\mu)/T\Big]}s
\end{equation}
for the non-relativistic fermion. 
Such a distribution function is also used for describing the spectrum of raindrop size: during the crossover phase transition between gas and liquid, the formation of a raindrop (water vapor condenses into rain droplets, rain droplets evaporation, rain droplets collide then merge or smash).

The number density of strangeon nuggets is given by
\begin{eqnarray}
n_{S}=\int_{D_c}^\infty n_0 e^{-D/R_c} dD,
\label{eq:nbnac}
\end{eqnarray}

The nuggets with $A>A_c$ form the stable strangeon nuggets that are non-relativistic, so they should follow the classical Maxwell-Boltzmann velocity distribution. Then similar to the ideal gas, the EOS reads
\begin{eqnarray}
p &=& {1\over 3} \int_{D_c}^\infty dD \ n_0 e^{-D/R_c} m_S(D)v(D)^2_{rms}\nonumber\\ 
 &=&{1\over 3} \int_{D_c}^\infty dD \ n_0 e^{-D/R_c} m_S(D) 4\pi\Big[{m_S(D) \over 2\pi T}\Big]^{3/2}\int_0^\infty dv\  v^2  exp{\Big[{-m_S(D) v^2 \over 2T}\Big]}\nonumber\\ 
&= &n_S T.
\end{eqnarray}

The contribution of the antiparticle to the thermodynamic quantities should have similar relations as Eqs.~(\ref{eq:eoshe}-\ref{eq:eoshs}). Therefore, the thermodynamic quantities, i.e., the pressure density, the entropy density, the energy density, and the Helmholtz free energy per baryon are:
\begin{eqnarray}
p_S&=& n_{S\pm}T, \label{eq:eosacp}\\
s_S&=&\int dD n_{S\pm}(D)\left\{{\ln}\left[\frac{\left({2\pi m_{S}(D) T}\right)^{3/2}}{n_{S\pm}(D)}\right]+\frac{5}{2}\right\},  \\
\epsilon_S&=&\frac{3}{2}n_{S\pm}T, 
\label{eq:eosace}\\
f_S&=&\epsilon_{S\pm}-Ts_{S\pm},
\label{eq:eosacf}
\end{eqnarray}
$n_{S\pm}$ is the number density of both positive particles $n_{S+}$ and negative particles $n_{S-}$.
$\epsilon_S$ is the average kinetic energy density, without the contribution of the rest mass. 
Because of the large number of $A_c$, the number density of strangeon nugget is very small (compared with the number density of ordinary hadrons) even with the contribution of particle-antiparticle pairs, which is about no larger than ten times the net number density. 
Energy release (i.e., entropy decrease) during the strangeon nugget formation could explain this, and the released energy could transfer to hadrons (entropy increase). The energy of HS phase is inherited from quark energy. Therefore, the energy density (and other thermodynamic quantities) of the HS phase is almost independent of the fraction of strangeon nuggets. As a result, all these thermodynamic quantities of strangeon nuggets are negligible compared with those of hadrons. Strangeon nuggets should satisfy the chemical potential equilibrium:
\begin{eqnarray}
\mu_{S}=A\left(\mu_u+\mu_d+\mu_s\right)=A\mu_\Lambda=A\mu_n.
\label{eq:mus}
\end{eqnarray}

The total entropy density of hadron and strangeon nuggets should have ($\frac{s}{n_b}$ is a constant)
\begin{eqnarray}
s_{HS}=s_H+s_S=n_{bHS}\frac{s}{n_b}.
\end{eqnarray}
Moreover, since $D= D_0A^{1/3}$, the baryon density of strangeon nugget under this distribution is given by
\begin{eqnarray}
\rho_{S}=\int_{D_c}^\infty n_0 \Big({D\over D_0}\Big)^3 e^{-D/R_c} dD,
\end{eqnarray}
then we have the mass density of the ordinary baryons (i.e., $A<A_c$ component ):
\begin{equation}
\label{eq:frac}
  f_{baryon} = {\int_0^{D_c} D^3 e^{-D/R_c} dD \over \int_0^{\infty} D^3 e^{-D/R_c} dD} = 0.1429. 
\end{equation}
Therefore, the stable strangeon nuggets constitute $\sim 85\%$ of the total baryon mass density, which could be an explanation of the dark matter without introducing the exotic theory. It is worth mentioning that such a case is based on the assumption that the stable nuggets are completely free from the other cosmological constraints. For example, if one only consider the geometry cross-section of the strangeon nuggets, i.e., $\sigma_{abs} = \pi R^2$, then the absorption rate of the neutron (For large size nuggets, protons are repelled by the surface electrostatic potential) is given by $dn_n/dt= \sigma_nv_n n_S \propto 1/R = 1/A^{1/3} $. Our previous study \cite{Lai:2010JCAP} showed that if the nuggets have a uniform size, then with baryon number $A>10^{25}$ , they are completely free from the primordial neutron-to-proton ratio. However, the relatively small nuggets could form bound state with light nuclei during nucleosynthesis epoch. Therefore, a detailed Big Bang Nucleosynthesis network is necessary to provide the realistic constraints on the distribution function of the number density of strangeon nuggets, this is out of the range of discussion in the present work.

\subsection{Crossover Phase Transition}
\label{sec:crosspt}
Similar to the idea in Ref.~\cite{Masuda2013a, Masuda2013b,Masuda2016a,Masuda2016b},
we define the crossover region (also known as the three-window modeling) around the chemical freeze-out temperature $T_c$ $\sim$ 170 MeV with $T_c - \Gamma<T<T_c + \Gamma$, where $\Gamma$ represents the temperature range of the QCD phase transition. The EOS of both the quark phase and the HS phase are described in Sec. \ref{sec:Qphase} and Sec. \ref{sec:Hphase} for $T\gg T+\Gamma$ and $T\ll T-\Gamma$, respectively.
It should be noted that the usually mentioned crossover region refers to that hadrons are hybrid with quarks. They coexist and interact strongly~\cite{Masuda2013a, Masuda2013b,Masuda2016a, Masuda2016b, Kojo2016, Whittenbury2016, Li2018}.
However, the crossover region we discussed is an entirely different one for the constitute:
in the crossover region, quarks collided and were confined in different-sized strangeon nuggets with baryon number $A$, which is one-third of the quark number. At the end of the crossover phase transition, all nuggets with $A< A_c$ have been destroyed and formed the nucleons. The $A> A_c$  components could survive. The realistic EOS of the crossover phase transition requires a detailed evaporation and interaction mechanisms of the strangeon nuggets, including evaporation productions and decay timescale for different cluster sizes. We do not consider the exact mechanism in this work.

We perform a smooth interpolation of the Helmholtz free energy per baryon $f$ between the HS phase and the QGP phase:
\begin{eqnarray}
f_\text{C}\left(T; \frac{s}{n_b}\right)=f_\text{Q}\left(T; \frac{s}{n_b}\right)\chi_{+}+f_\text{HS}\left(T; \frac{s}{n_b}\right)\chi_{-}
\end{eqnarray}
with $\chi_{+}$ and $\chi_{-}= 1-\chi_{+}$ are the weight functions. The Helmholtz free energy per baryon $f$ should be the function of number density $n_b$ and temperature $T$ (or entropy $S$) under $\beta$ equilibrium. Since the Universe evolves with $T$ decrease as well as a fixed value of $s/n_b$, $n_b$ becomes a function of $T$. So we assume $\chi_{+}$ depends on $T$ :
\begin{eqnarray}
\chi_{\pm}=\frac{1}{2}\left[1\pm \text{tanh} \left(\frac{T-T_c}{\Gamma}\right)\right]
\end{eqnarray}

The baryon number conservation reads
\begin{eqnarray}
n_b&=&\frac{1}{3}\left(n_u+n_d+n_s\right)\chi_{+} +\left(n_p+n_n+n_{A_i}\right)\chi_{-} .
\end{eqnarray}
The thermal quantities are
\begin{eqnarray}
\label{eq:threewindow}
s_\text{C}\left(T; \frac{s}{n_b}\right)=s_\text{Q}\left(T; \frac{s}{n_b}\right)\chi_{+}+s_\text{HS}\left(T; \frac{s}{n_b}\right)\chi_{-},\\
\epsilon_\text{C}\left(T;\frac{s}{n_b}\right)=\epsilon_\text{Q}\left(T; \frac{s}{n_b}\right)\chi_{+}
+\epsilon_\text{HS}\left(T; \frac{s}{n_b}\right)\chi_{-},\\
P_\text{C}\left(T;\frac{s}{n_b}\right)=P_\text{Q}\left(T; \frac{s}{n_b}\right)\chi_{+}
+P_\text{HS}\left(T; \frac{s}{n_b}\right)\chi_{-}.
\end{eqnarray}
A smooth transition ($P$ and $\epsilon$) is necessary so there is no latent heat like the first-order one. 


\section{Results}
\label{sec:results}

\begin{figure*}[htb]
\includegraphics[bb=0 5 580 400, width=7.8
cm,clip]{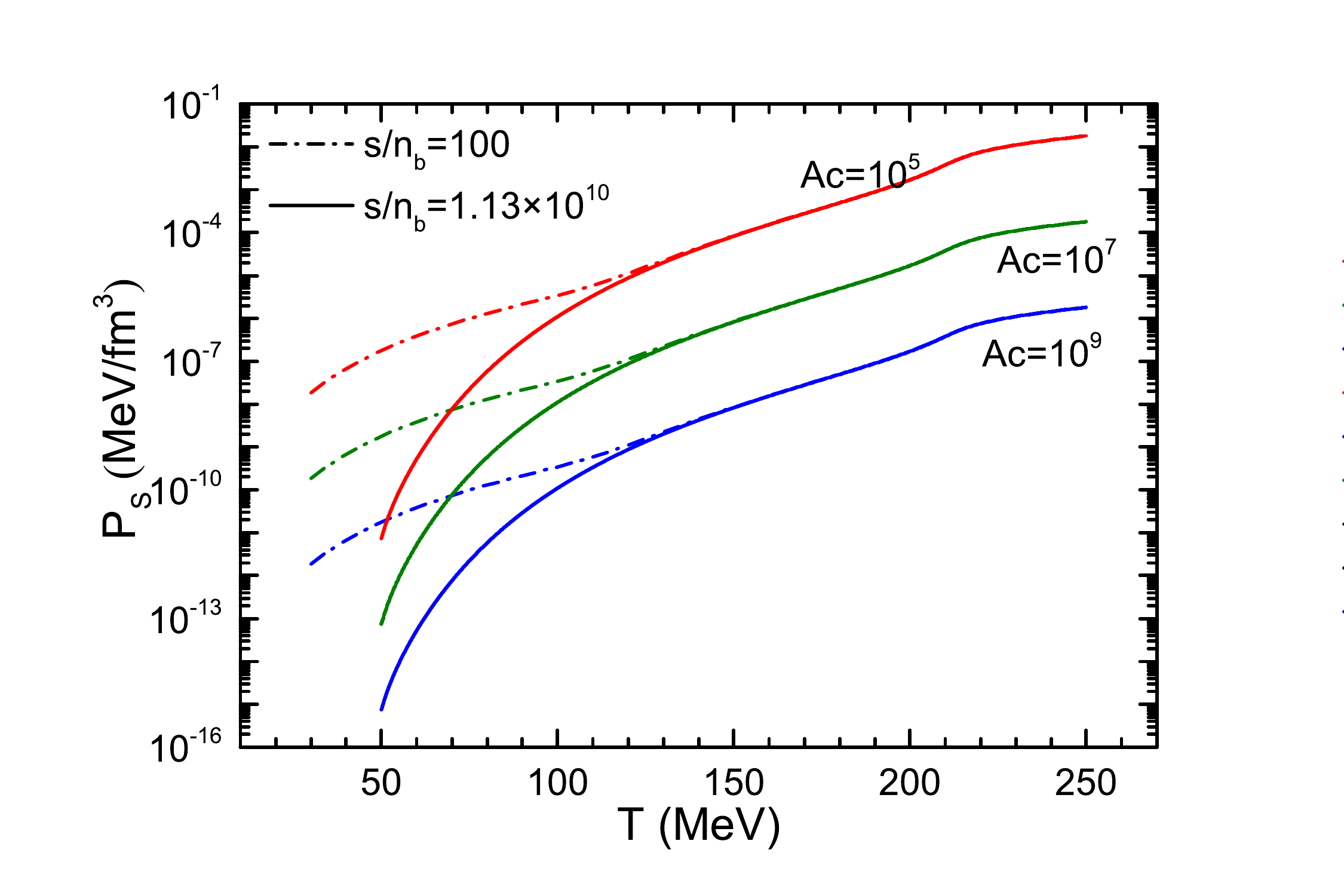}%
\includegraphics[bb=0 5 580 400, width=7.8
cm,clip]{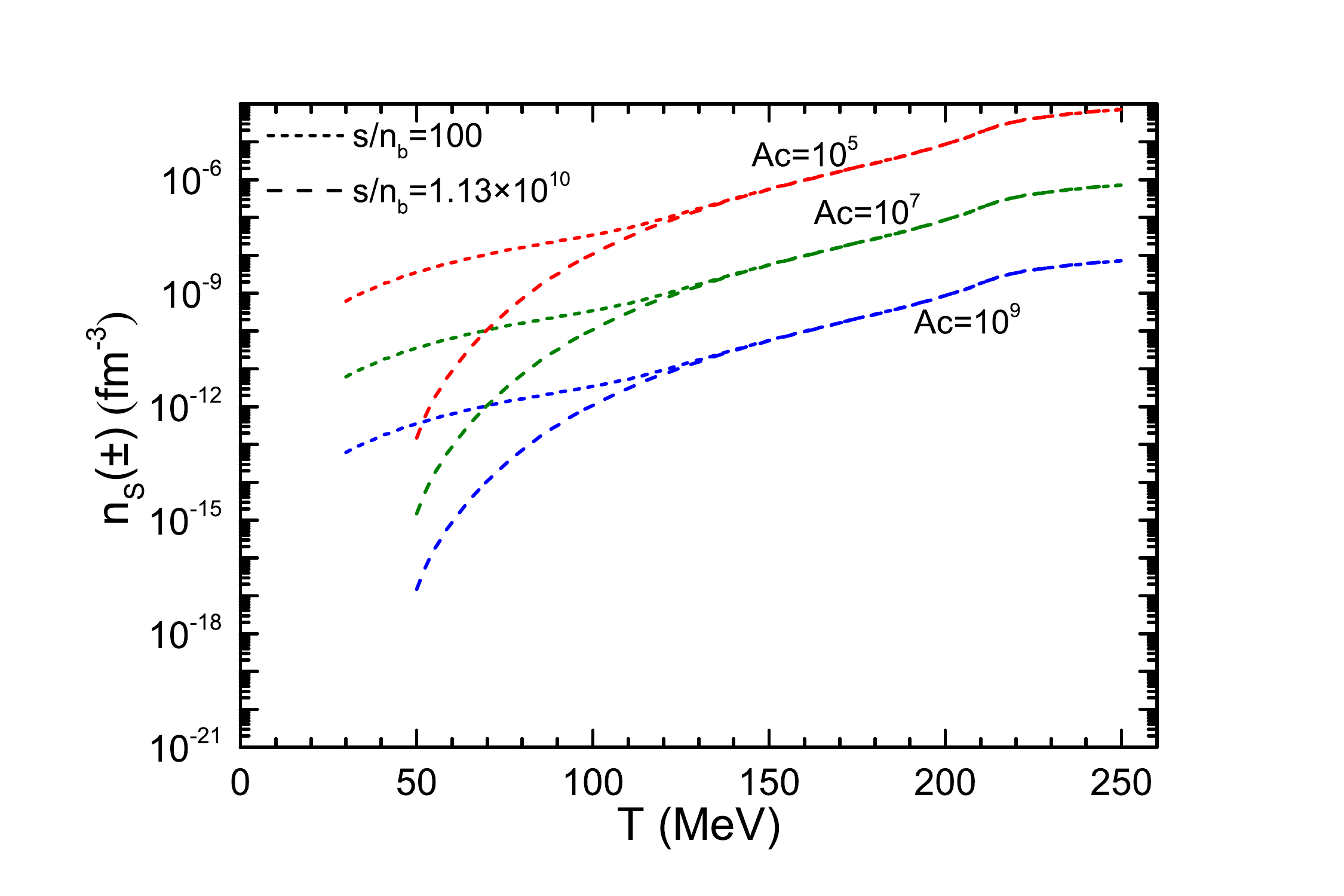}%
\caption{The left panel shows the pressure $P_S$ of strangeon nugget as functions of temperature $T$ and the right panel shows the total number density (i.e., the sum of particles and antiparticles) as functions of $T$. We choose three values of the critical baryon number $A_c = 10^5,10^7,10^9$ plotted in red, blue, and green, separately. The high and low entropy cases are shown in dash-dotted lines and solid lines, respectively.}
\label{fig:2eosac}
\end{figure*}
\begin{figure*}[htb]
\includegraphics[bb=5 5 580 400, width=7.8
cm,clip]{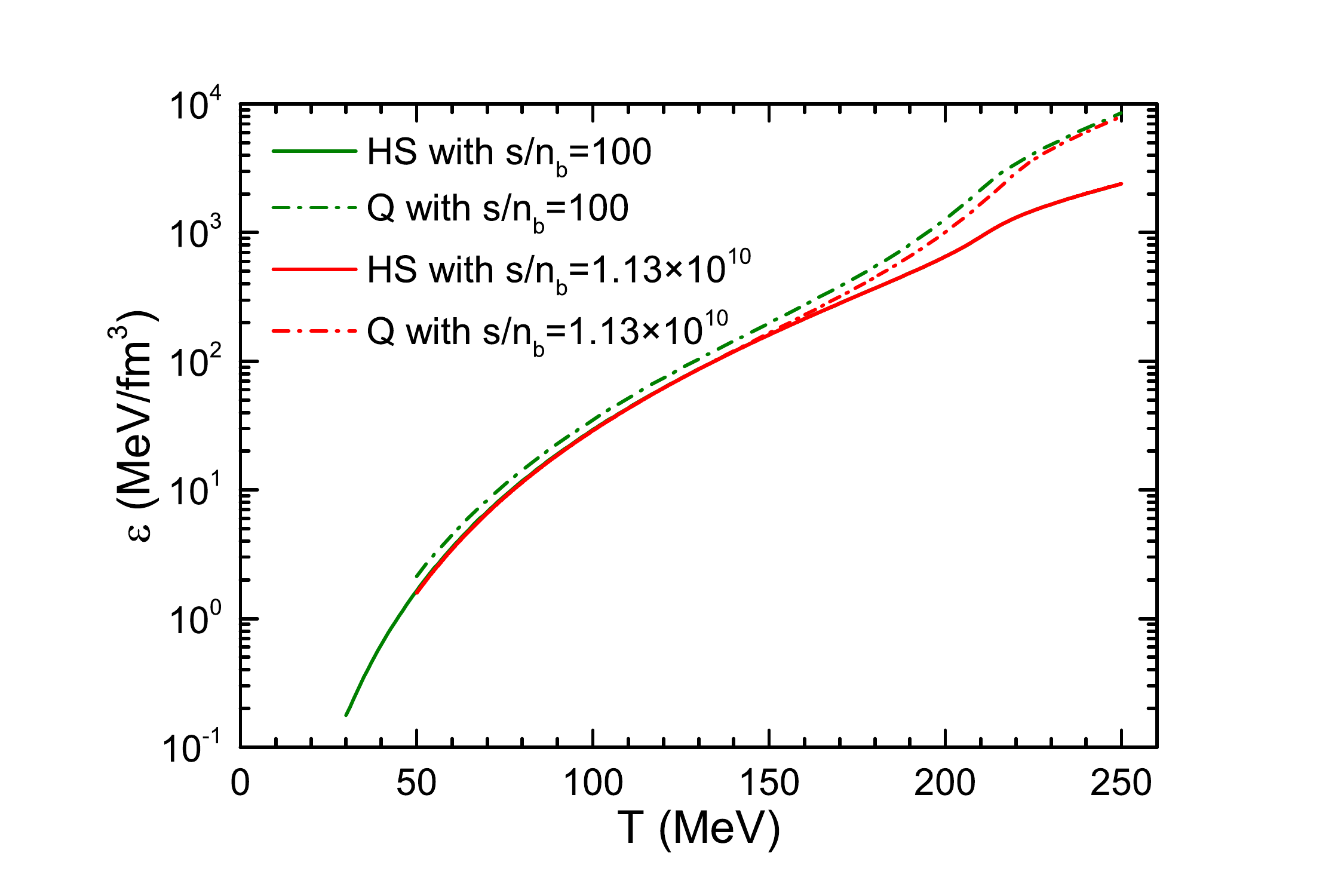} 
\includegraphics[bb=5 5 580 400, width=7.8
cm,clip]{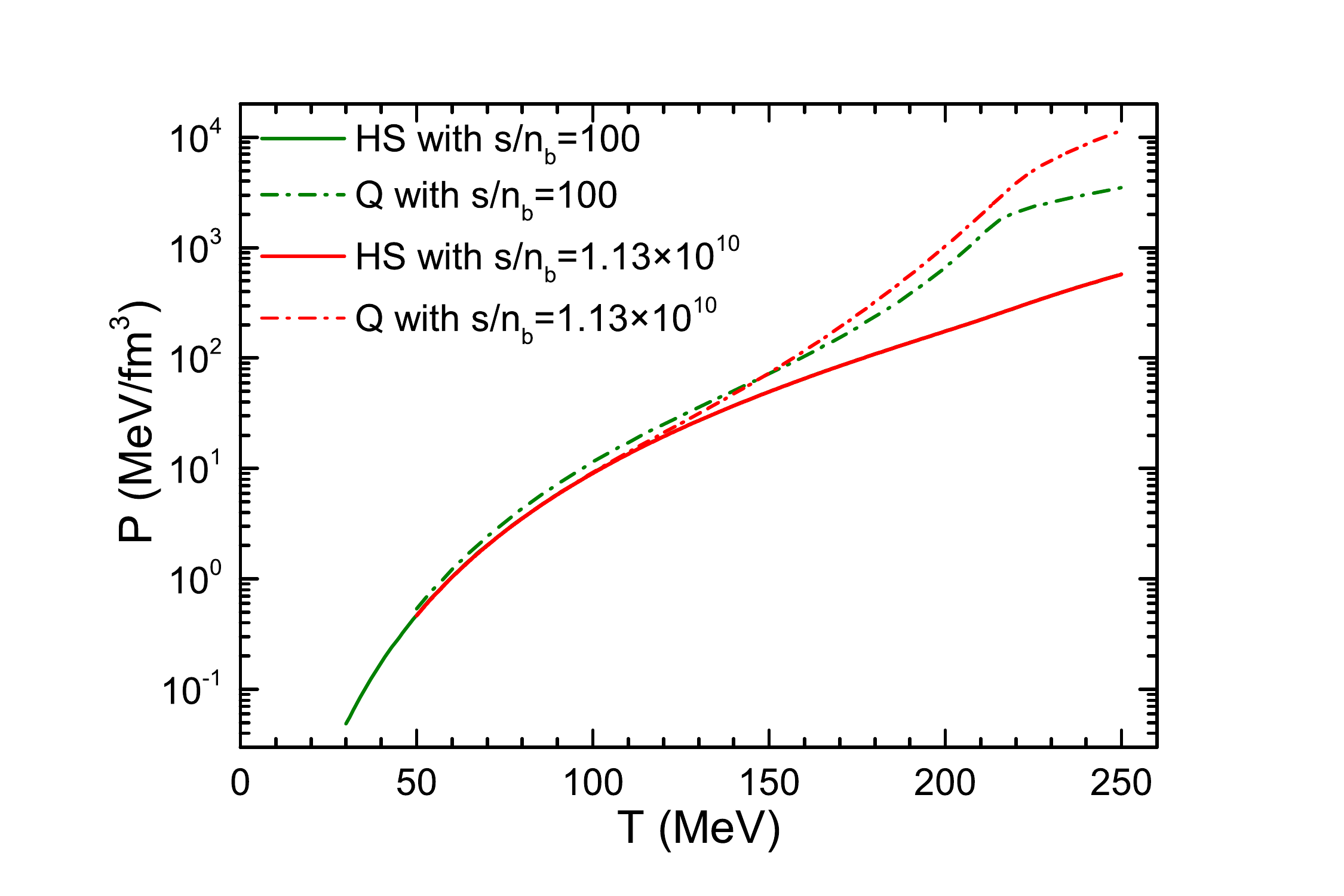} \\
\includegraphics[bb=5 5 580 400, width=7.8
cm,clip]{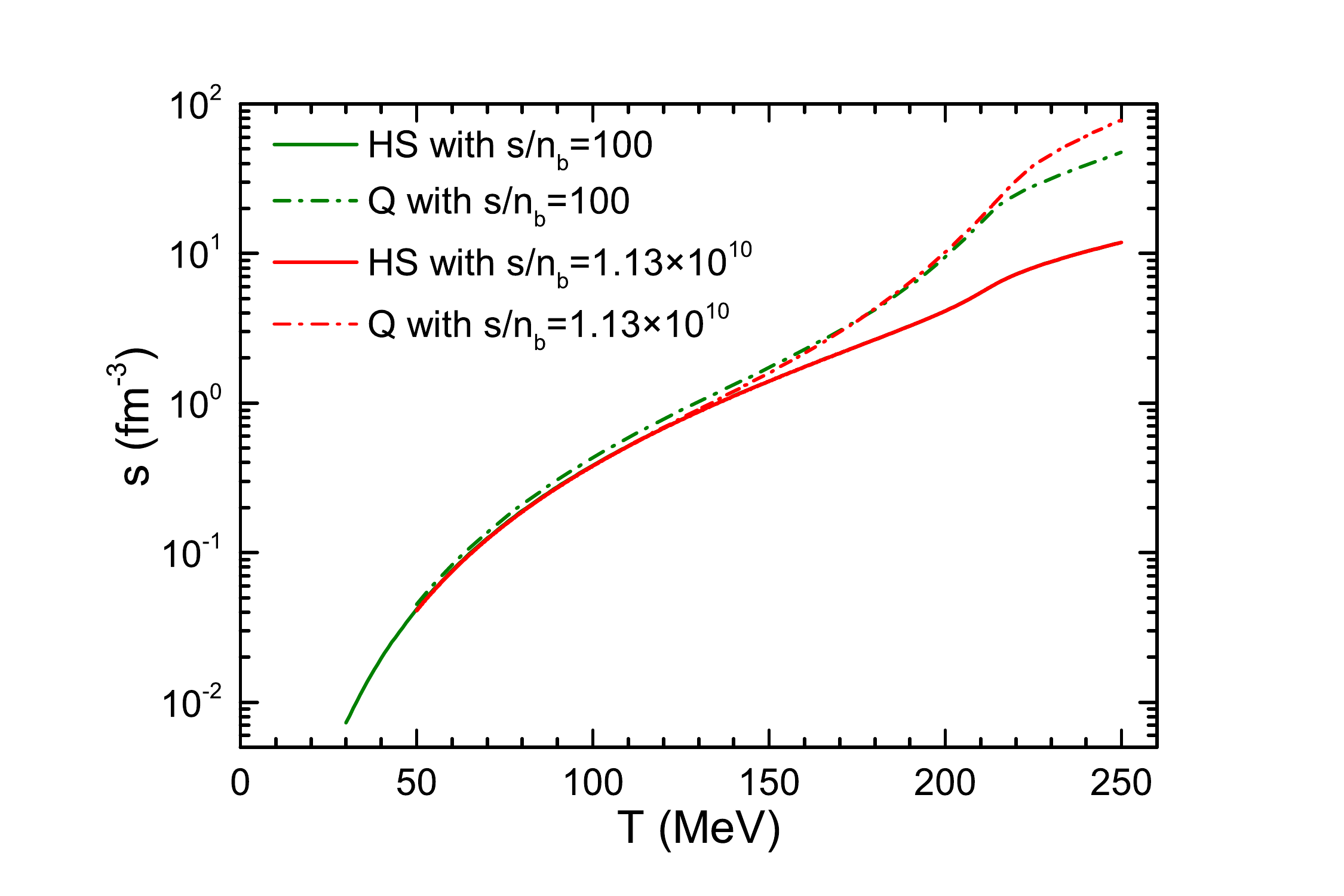}
\includegraphics[bb=5 5 580 400, width=7.8
cm,clip]{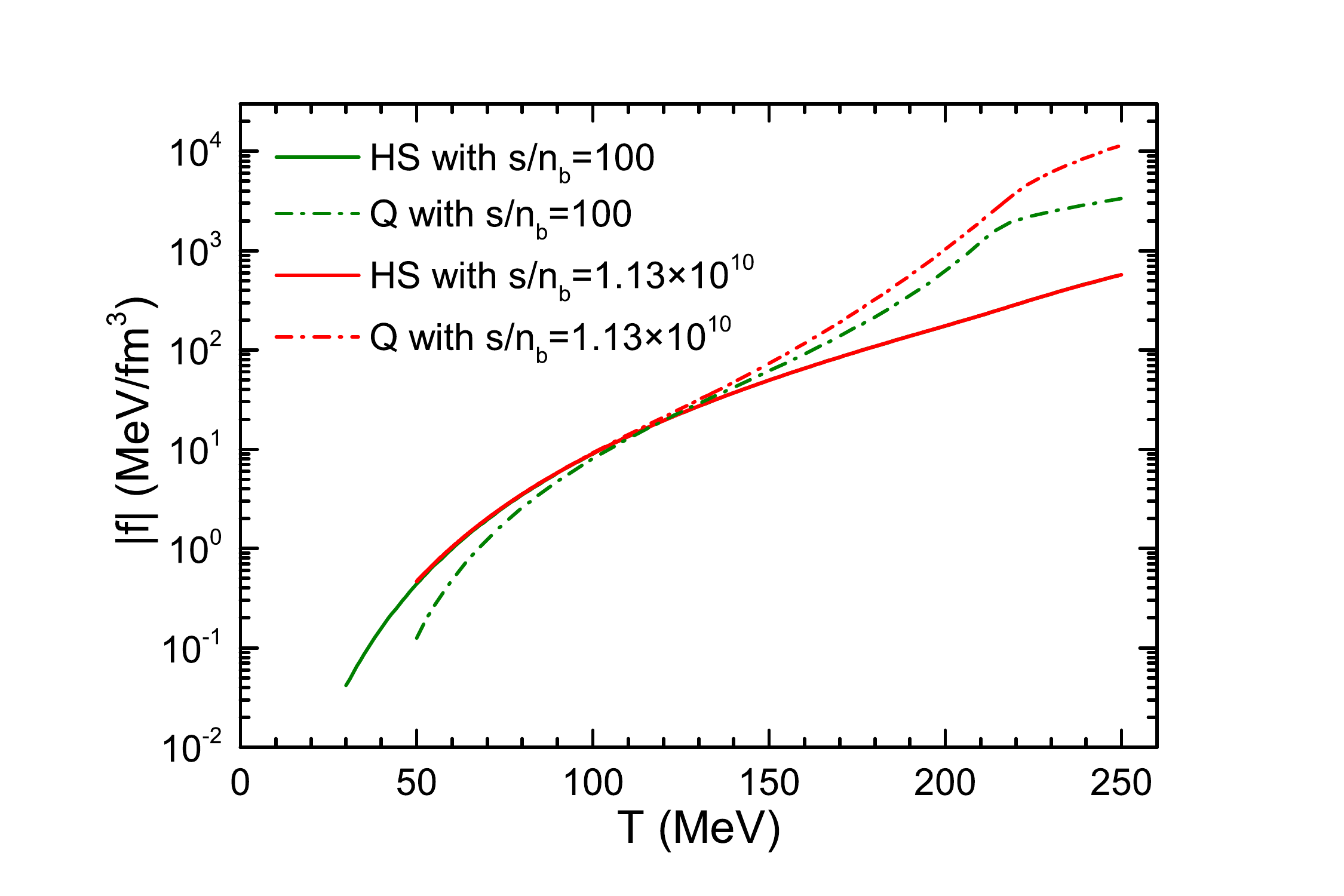}%
\caption{The energy density $\epsilon$,  pressure $P$, entropy density $s$, and free energy density $f$ for HS phase and quark phase as a function of temperature $T$. In each panel, the QGP phase (Q) and the HS phase results are presented in dash-dotted lines and solid lines, respectively. The high and low entropy cases are shown in red and green. Also, notice that the free energy density $f$  is a negative quantity, so it is plotted in its absolute value on the right bottom panel.}
\label{fig:3eos}
\end{figure*}


In Fig.~\ref{fig:2eosac} we show the pressure $P_S$ (left panel) and the number density $n_S$ of strangeon nuggets as functions of the cosmic temperature $T$. We choose different values of the critical baryon number $A_c$ as $A_c= 10^5, 10^7, 10^9$ showing in red, blue, and green lines, respectively. For both panels, we present both low and high entropy cases with $s/n_b = 100$  (dash-dotted lines) and $s/n_b = 1.13\times 10^{10}$ (solid lines) separately. The pressure comes from the motion of both particles and antiparticles so higher temperature may have higher pressure due to the co-existence of particles and antiparticles. Also, the value of $n_S(\pm)$ represents the total number density of both particles and antiparticles, so the high and low entropy cases are undistinguished at a higher temperature. When $T$ decreases to about $120~\MeV$, almost all the antiparticles annihilate with their corresponding particles, so the pressure decreases. For a larger value of $A_c$, the formed strangeon nuggets contain more baryon, so the number density $n_S(\pm)$ becomes smaller (right panel) and vice versa. A similar trend also can be seen on the left panel. We also expect that other thermodynamic quantities such as $\epsilon_S$, $f_{S}$, and $s_S$ have the same magnitude as $P_S$ since they are all related to the EOS as we shown in Eq. (\ref{eq:eosace}). As a result, strangeon nuggets can be treated as clusters of baryonic matter, and their contribution to the thermodynamics of the early Universe is negligible. This is one of the reasons that strangeon nugget is a potential candidate for dark matter.

\begin{figure*}[htb]
\centering
\includegraphics[bb=5 5 580 400, width=7.8
cm,clip]{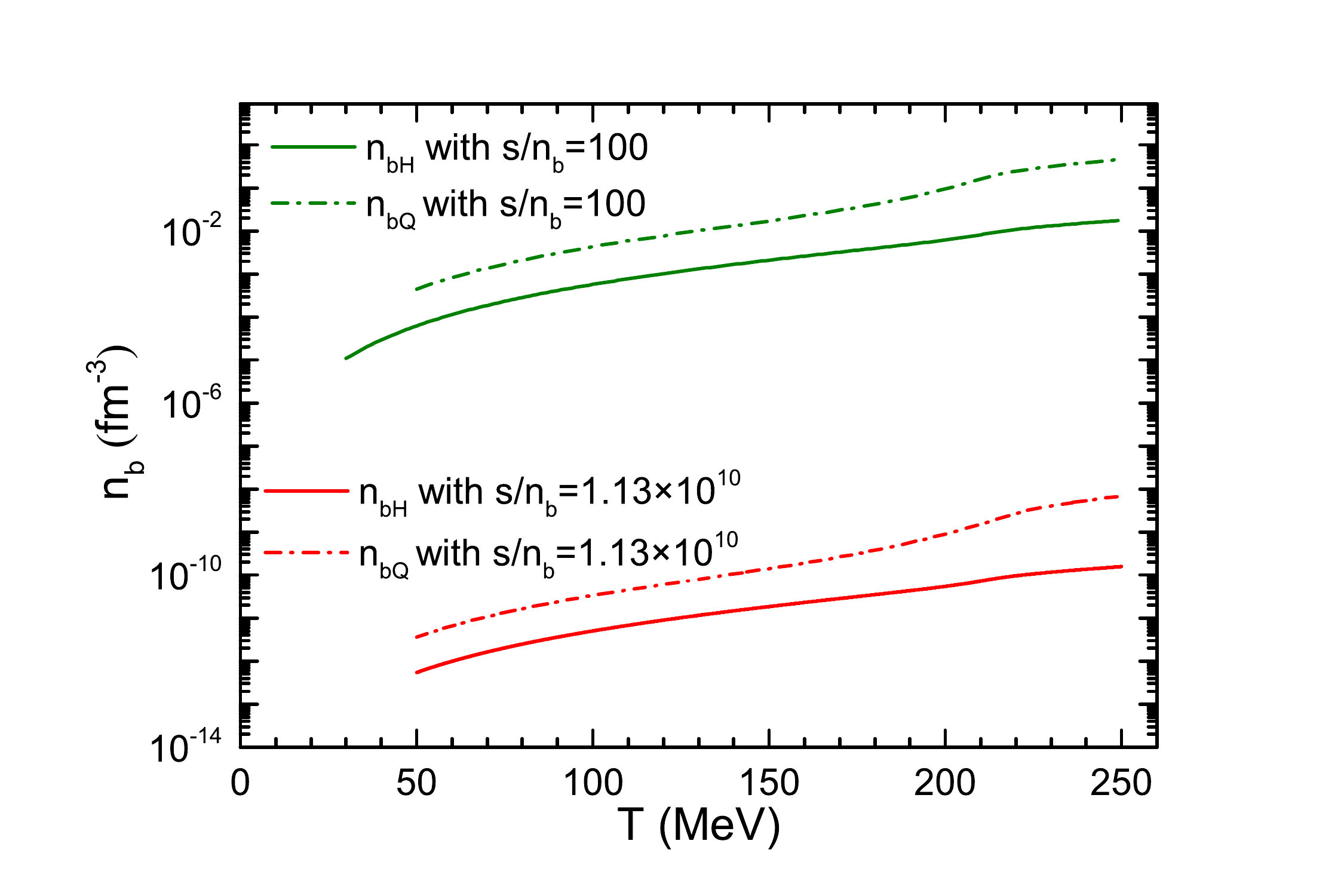}%
\includegraphics[bb=5 5 580 400, width=7.8
cm,clip]{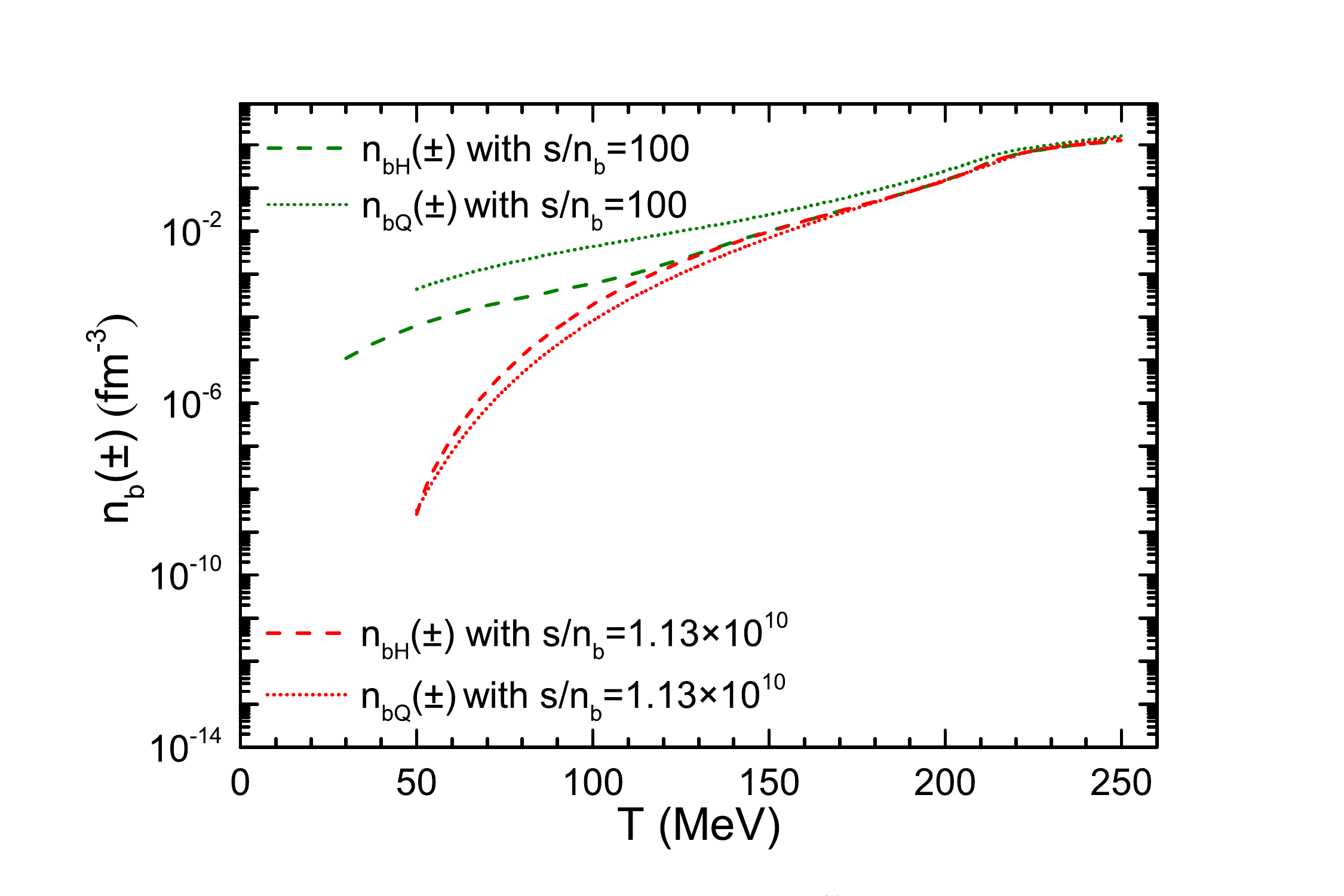}
\caption{The left panel shows the net baryon number density $n_b$ for both hadrons and quarks as functions of $T$, the right panel shows the same relation but for the total baryon number density $n_b(\pm)$. We also present both the low entropy (plotted in green) and high entropy cases (plotted in red) for comparison on each panel.}
\label{fig:4tn}
\end{figure*}

\begin{figure*}[btp]
\centering
\includegraphics[bb=5 5 580 400, width=5.2
cm,clip]{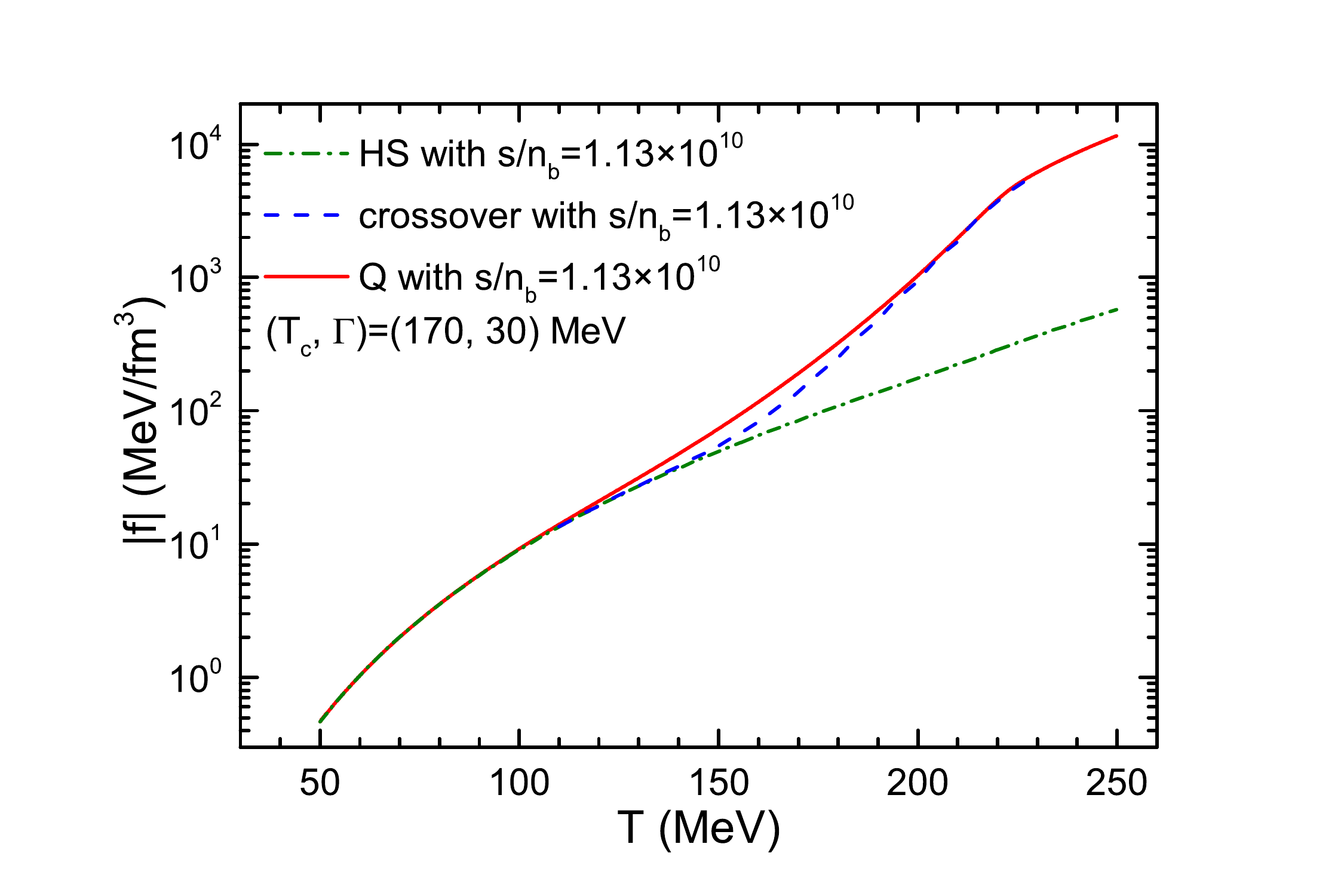}%
\includegraphics[bb=5 5 580 400, width=5.2
cm,clip]{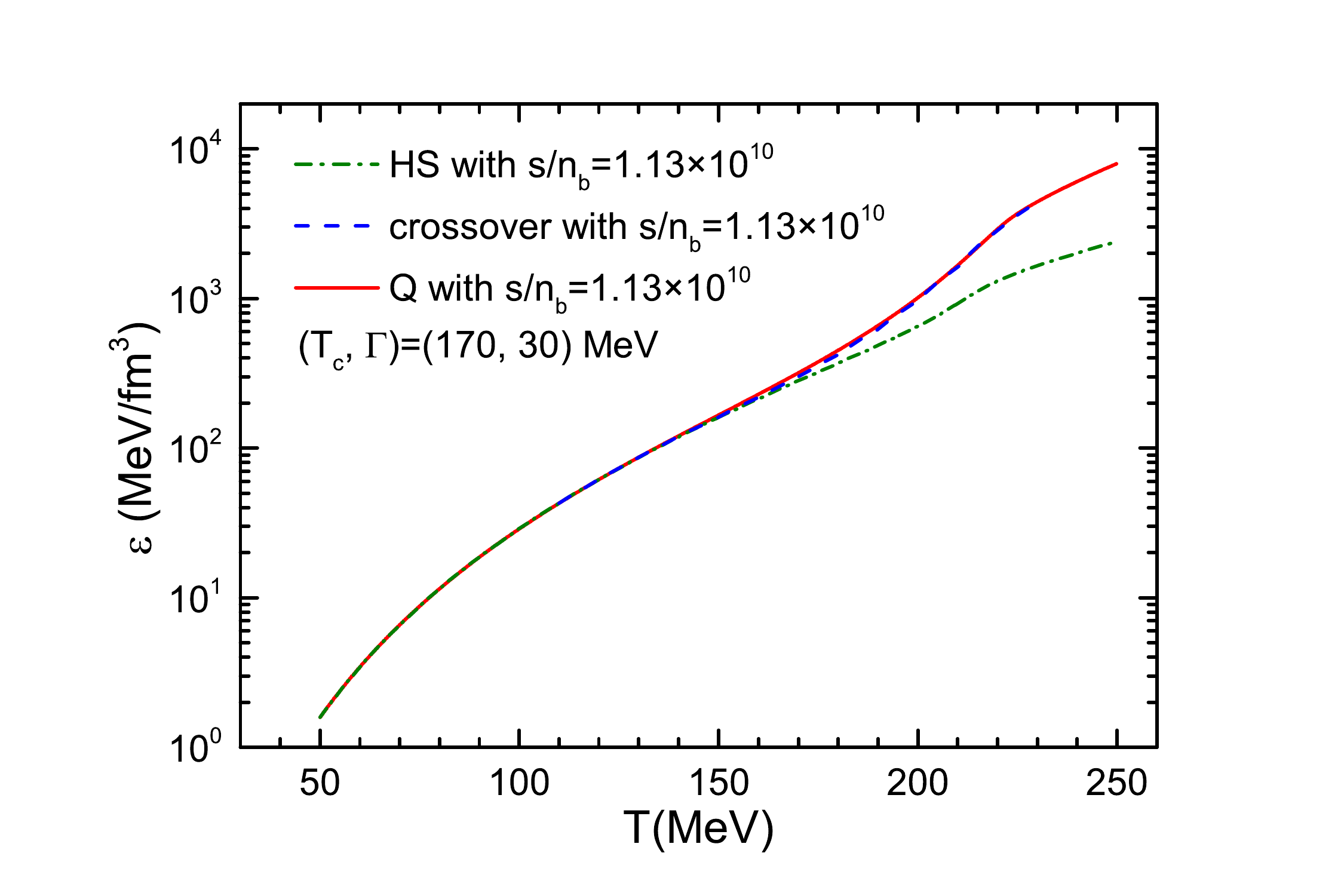}%
\includegraphics[bb=5 5 580 400, width=5.2
cm,clip]{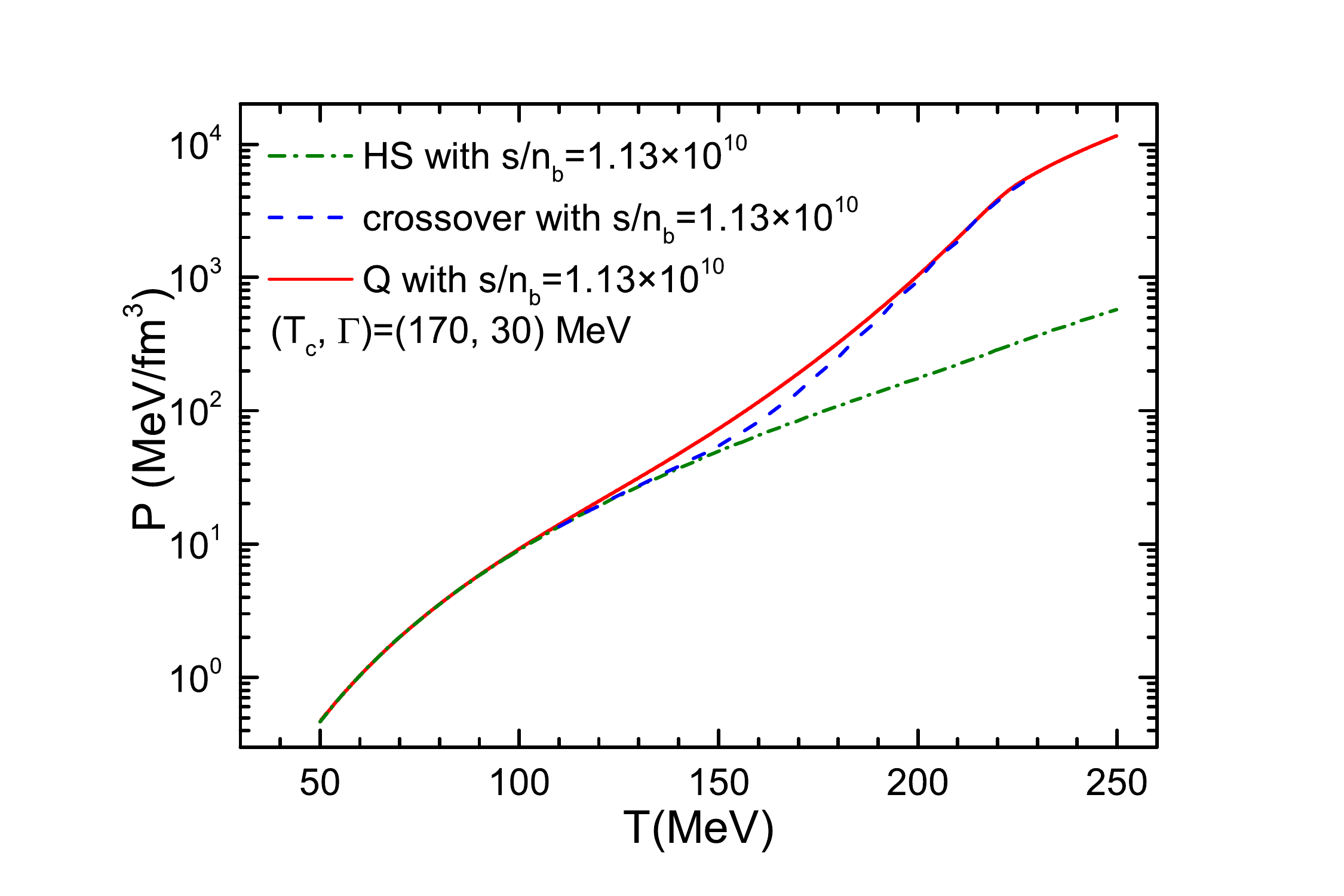}%
\caption{The free energy density $f$, energy density $\epsilon$ and pressure $P$ as a function of temperature $T$ for quark phase (solid red line), crossover region (blue dash line), and HS phase (dash-dot green line). The crossover phase transition occurs at $(T_c, \Gamma)=(170, 30)$ MeV, i.e., the blue lines are the interpolation between the red and green lines in this temperature range.}
\label{fig:5eoscross}
\end{figure*}

Fig.~\ref{fig:3eos} shows the thermodynamic quantities ($\epsilon$, $P$, $s$, and $f$) of HS phase  and quark phase as a function of temperature $T$, respectively. A rapid increase of $\epsilon$,  $P$, and $s$ with temperature can be seen above $T = 150~\MeV$ because the early Universe contains particles and antiparticles for high temperature, and their number density is proportional to $T^3$. Also, for a high temperature, the main component of the elementary particles is free quarks (dash-dotted lines). The statistical equilibrium determines their number density, and the equilibrium number density drops quickly at a lower temperature. Below $T= 150$ MeV, they become the hadrons (see Fig.~\ref{fig:4tn} for more details). Moreover, at a low temperature, the entropy density of quark phase (dash-dotted lines in the left-bottom panel) is in concordance with the entropy density of HS phase (solid lines in the left-bottom panel), this could be explained by  the mass fraction of strangeon nuggets: in our model, the mass fraction of those nuggets is  $0.8549$ to the total baryon, so considering all the quark turns into the hadrons at low temperature, the number ratio between HS and  quarks is $n_{HS} \sim 0.85/0.15 n_H \sim n_Q$. The Helmholtz free energy $f$ behaves opposite to the other thermodynamic quantities (notice that in the right bottom panel, $f$ is plotted in absolute value). The number of particle species coupled with the plasma decreases with temperature, and so does the total amount of energy available. Such an effect could explain the decreasing trend of the Helmholtz free energy since it describes the difference between internal energy and heat. It is also worth mentioning that since these thermodynamic quantities should evolve continuously, as we describe in the previous section, the phase transition should occur when the interpolation of $f$, $\epsilon$and $P$ could connect two phases (see Fig. \ref{fig:5eoscross} for details).

Fig. \ref{fig:4tn} describes the net number density $n_b$ (left panel) and the total baryon number density $n_b{\pm}$  (right panel) as a function of temperature $T$. The net number density of both phases is shown in dash-dotted lines for the QGP phase and solid line for the HS phase, respectively. The total number density is shown in the dotted line and dashed line, respectively. The Universe has a conserved total number density of baryon $N_b$ so that a larger $s/n_b$ means that with a given $T$, $s$ contributes more energy. A smaller value of $s/n_b$ corresponds to a denser Universe. The $n_b$ is fixed so that the ratio between net number density $n_b$ and total number density $n_b\pm$ is preserved. At a high temperature, mainly above $100$ MeV, there is a significant gap between $n_b$ and $n_{b\pm}$. At a low temperature, almost all the antiparticles annihilate with its mirror so that $n_{\pm}$ is similar to $n$.

\begin{figure}[htb]
\centering
\includegraphics[bb=40 5 560 400, width=8
cm,clip]{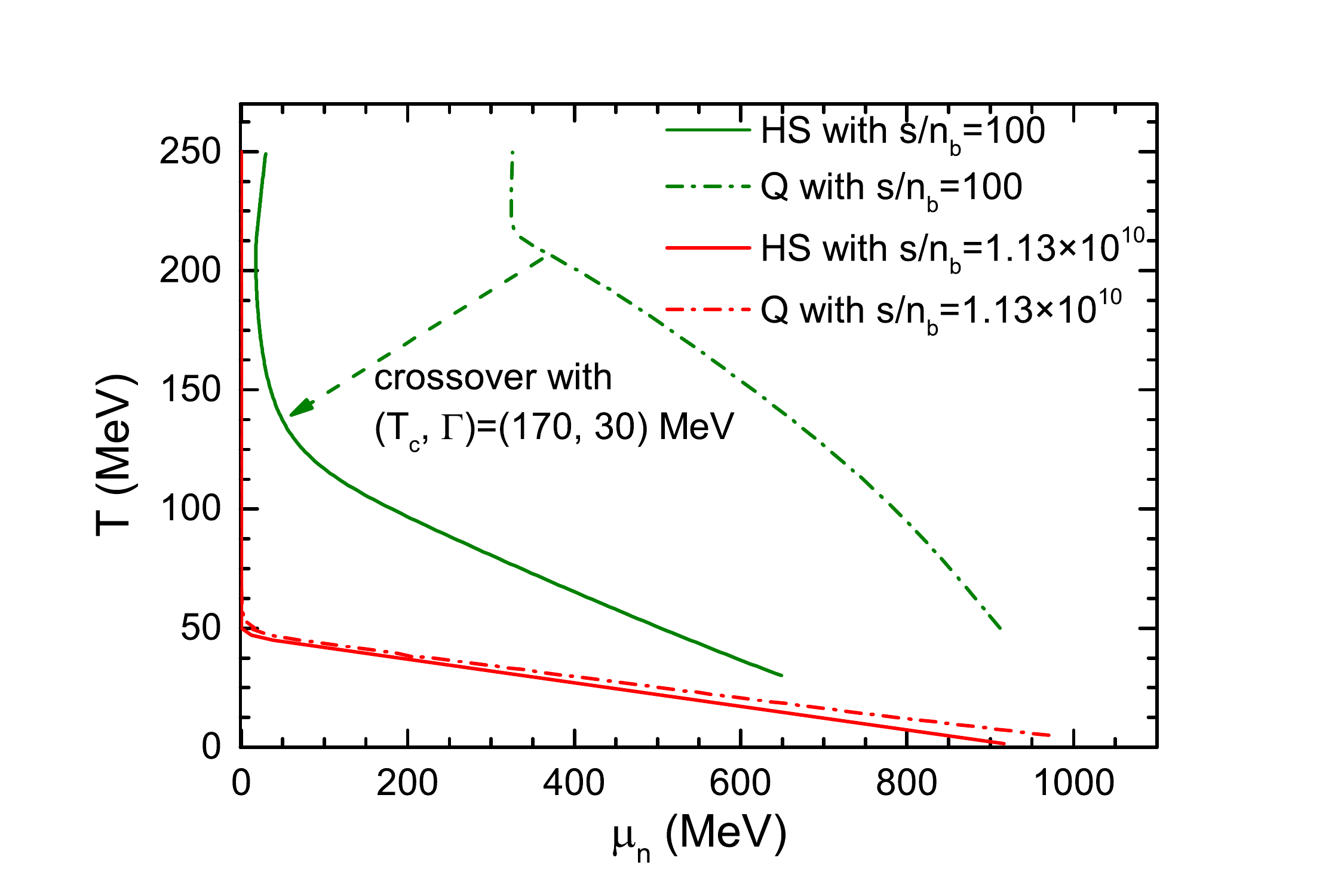}%
\caption{The QCD phase diagram. Dash-dot and solid lines represent the evolutionary trajectory of the QCD phase and HS phase, respectively. The separation between the two phases for the $s/n_b = 1.13\times 10^{10}$ case is too tiny in this plot, so we use the dashed arrow to illustrate the crossover phase transition for the case of $s/n_b = 100$.}
\label{fig:6qcd}
\end{figure}

The above discussion of the thermodynamic quantities in both the QGP and HS phase are the groundwork for the crossover phase transition. As we mentioned in Sec. \ref{sec:crosspt}, all the thermodynamic quantities should change continuously with cosmic temperature during the crossover phase transition. The transition relation is taken from Eq. (\ref{eq:threewindow}). In Fig. \ref{fig:5eoscross}, we show the three-window relation between the thermodynamic quantities ($f$, $\epsilon$ and $P$) and cosmic temperature $T$. The three windows are the QGP phase (solid lines), narrow transition window (dashed lines), and HS phase (dash-dotted lines). The narrow window of the crossover is set as ($T_c, \Gamma$)=(170, 30) MeV, and a rapid decrease of energy density $\epsilon$ and pressure $P$ is observed in this region. Such a trend corresponds to particle-antiparticle pairs annihilation. Finally, we show the QCD phase diagram in Fig. \ref{fig:6qcd}. The horizontal axis is the chemical potential of the neutron. This value is converted from the chemical potential of quark: $\mu_n=2\mu_d+\mu_u$. This value change can represent the QCD transition trajectory in the QCD phase diagram since the hadrons is still in the statistical equilibrium at $T>10$ MeV. In this figure, the curves represent the evolutionary trajectories of both the QCD phase (dash-dotted lines) to the HS phase (solid lines). The chemical potential $\mu_n$ is correlated to the baryon number density $n_b$, so a smaller $s/n_b$ value could shift the trajectory to the right side in this figure. The crossover phase transition occurred at $(T_c,\Gamma)=(170,30)$ MeV, the arrow illustrates such a transition for $s/n_b=100$.

\section{Summary}
\label{sec:summary}
This work investigated a crossover QCD phase transition in the early Universe. At a high cosmic temperature, three flavors of quarks (i.e., $u$, $d$, and $s$) exist. Then during the crossover phase transition, these quarks could collide and nucleate to form the strangeon nuggets. The small nuggets will decay quickly to the hadrons, while the large nuggets will become stable in this scenario, surviving from the early Universe.

We consider such a crossover phase transition occurred at a temperature $T\sim 170$ MeV. The SU(3) PNJL model is used to describe the thermodynamic quantities Eq.(\ref{eq:pnjlp}) - Eq.(\ref{eq:pnjls}) for the quark phase at high temperature. The evaluation of the thermodynamic quantities for the hadron phase at low temperature is based on the RMF model Eq.(\ref{eq:eoshe} ) - Eq.(\ref{eq:eoshs}), as shown in Fig. \ref{fig:3eos}. The crossover phase transition is a smooth transition between these two phases. The three-window model is used in our work, i.e., the smooth interpolations of the Helmholtz free energy per baryon $f$, the pressure $P$ and the energy density $\epsilon$ between quark and hadron phases (Fig. \ref{fig:5eoscross}).

After temperature decreases under $100$ MeV, the stable strangeon nuggets could exist with different baryon number $A$. Since heavier quark clusters are more difficult to form, an exponential distribution function with a critical parameter $A_c$ is introduced to describe the number density of the strangeon nuggets Eq. (\ref{eq:sndis}). The nuggets with the baryon number $A>A_c$ are supposed to be stable.
Certainly, the critical baryon number $A_c$, determined by both the weak and strong interactions,  should be quantitatively approached by QCD-based calculations in the future.
Nonetheless, we chose its value as $10^{5,7,9}$ based on the previous estimations \cite{Wang2018}.
Due to the large baryon number of the strangeon nuggets, we use the non-relativistic EOS to calculate their thermodynamic quantities Eq.(\ref{eq:eosacp}) - Eq.(\ref{eq:eosacf}). 
Although a detailed primordial nucleosynthesis study is necessary to provide a realistic constraint on such a strangeon nugget model, the results show that the contribution of the stable strangeon nuggets to the total hadronic thermodynamics is negligible (Fig. \ref{fig:2eosac}). Moreover, the mass density of the strangeon nuggets to the total matter density is $\sim 0.85$. which indicates that the heavy strangeon nuggets are the potential candidate for cold dark matter.

\acknowledgments
We are grateful to Prof. Yuxin Liu and Prof. Motohiko Kusakabe for their helpful suggestion and comments. This work is supported by the National SKA Program of China (No. 2020SKA0120100).



\begin{thebibliography}{99}
\bibitem{Linde:1978px}
A.~D.~Linde,
Rept. Prog. Phys. \textbf{42}, 389 (1979)

\bibitem{Kibble:1980mv}
T.~W.~B.~Kibble,
Phys. Rept. \textbf{67}, 183 (1980)

\bibitem{Mazumdar:2018dfl}
A.~Mazumdar and G.~White,
Rept. Prog. Phys. \textbf{82}, no.7, 076901 (2019)
doi:10.1088/1361-6633/ab1f55
[arXiv:1811.01948 [hep-ph]].

\bibitem{Hindmarsh:2020hop}
M.~B.~Hindmarsh, M.~L\"uben, J.~Lumma and M.~Pauly,
SciPost Phys. Lect. Notes \textbf{24}, 1 (2021)
doi:10.21468/SciPostPhysLectNotes.24
[arXiv:2008.09136 [astro-ph.CO]].

\bibitem{Durrer:2013pga}
R.~Durrer and A.~Neronov,
Astron. Astrophys. Rev. \textbf{21}, 62 (2013)
doi:10.1007/s00159-013-0062-7
[arXiv:1303.7121 [astro-ph.CO]].

\bibitem{Kuzmin:1985mm}
V.~A.~Kuzmin, V.~A.~Rubakov and M.~E.~Shaposhnikov,
Phys. Lett. B \textbf{155}, 36 (1985)
doi:10.1016/0370-2693(85)91028-7

\bibitem{Farrar:1993sp}
G.~R.~Farrar and M.~E.~Shaposhnikov,
Phys. Rev. Lett. \textbf{70}, 2833-2836 (1993)
[erratum: Phys. Rev. Lett. \textbf{71}, 210 (1993)]
doi:10.1103/PhysRevLett.70.2833
[arXiv:hep-ph/9305274 [hep-ph]].

\bibitem{Kosowsky:1992rz}
A.~Kosowsky, M.~S.~Turner and R.~Watkins,
Phys. Rev. Lett. \textbf{69}, 2026-2029 (1992)
doi:10.1103/PhysRevLett.69.2026

\bibitem{Schettler:2010dp}
S.~Schettler, T.~Boeckel and J.~Schaffner-Bielich,
Phys. Rev. D \textbf{83}, 064030 (2011)
doi:10.1103/PhysRevD.83.064030
[arXiv:1010.4857 [astro-ph.CO]].

\bibitem{Boeckel:2010bey}
T.~Boeckel, S.~Schettler and J.~Schaffner-Bielich,
Prog. Part. Nucl. Phys. \textbf{66}, 266-270 (2011)
doi:10.1016/j.ppnp.2011.01.017
[arXiv:1012.3342 [astro-ph.CO]].


\bibitem{Cline:2008hr}
J.~M.~Cline, M.~Jarvinen and F.~Sannino,
Phys. Rev. D \textbf{78}, 075027 (2008)
doi:10.1103/PhysRevD.78.075027
[arXiv:0808.1512 [hep-ph]].


\bibitem{Weinberg:1974hy}
S.~Weinberg,
Phys. Rev. D \textbf{9}, 3357-3378 (1974)
doi:10.1103/PhysRevD.9.3357

\bibitem{Higgs:1964pj}
P.~W.~Higgs,
Phys. Rev. Lett. \textbf{13}, 508-509 (1964)
doi:10.1103/PhysRevLett.13.508

\bibitem{Witten1984} E. Witten,
Phys. Rev. D \textbf{30}, 272 (1984).

\bibitem{HotQCD:2014kol}
A.~Bazavov \textit{et al.},
Phys. Rev. D \textbf{90}, 094503 (2014).

\bibitem{Schmidt:2017bjt}
C.~Schmidt and S.~Sharma,
J. Phys. G \textbf{44}, no.10, 104002 (2017).

\bibitem{Aarts2023} G. Aarts, et al.,
arXiv:2301.04382.

\bibitem{Peebles2022} P. J. E. Peebles,
Annals of Physics \textbf{447}, 169159 (2022).

\bibitem{Lai:2010JCAP}
X.~Y.~Lai and R.~X.~Xu,
JCAP \textbf{05}, 028 (2010)

\bibitem{Kolb1990}
E. W. Kolb, and M. S. Turner,  
{\em The early universe}, Addison-Wesley Publishing Company (1990).
\bibitem{Schwarz2003} D. Schwarz, 
Ann. Phys. \textbf{12}, 220-270 (2003).

\bibitem{Aoki:2009sc}
Y.~Aoki, S.~Borsanyi, S.~Durr, Z.~Fodor, S.~D.~Katz, S.~Krieg and K.~K.~Szabo,
JHEP \textbf{06}, 088 (2009)

\bibitem{Xu2019}R. Xu,
AIP Conference Proceedings \textbf{2127}, Issue 1, 020014 (2019).

\bibitem{Lai2021}
X. Y. Lai, C. J. Xia, Y. W. Yu, and R. X. Xu, 
Res. Astron. Astrophys. \textbf{21}, 250 (2021).
%
\bibitem{Alcock1985} C. Alcock and E. Farhi, 
Phys. Rev. D \textbf{32}, 1273 (1985).
\bibitem{Madsen1986} J. Madsen, H. Heiselberg and K. Riisager, 
Phys. Rev. D \textbf{34}, 2947 (1986).
%
\bibitem{Sumiyoshi1991} K. Sumiyoshi and T. Kajino, 
Nucl. Phys. B (Proc. Suppl.) \textbf{24}, 80 (1991).
\bibitem{Bhattacharjee1993} P. Bhattacharjee, Jan-e Alam, B. Sinha, S. Raha,
Phys. Rev. D \textbf{48}, 4630 (1993).

\bibitem{Xu2003} R. X. Xu,
Astrophys. J, 596, \textbf{L59} (2003).

\bibitem{Lai2023} X. Lai, C. Xia and R. Xu, Advances in Physics X, 8, \textbf{2137433} (2023).

\bibitem{Xu2023}R. Xu, Astron.
Nachr., e20230008 (2023) arXiv:2212.10887.

\bibitem{Xu:1999bw}
R.~X.~Xu, G.~J.~Qiao and B.~Zhang,
Astrophys. J. Lett. \textbf{522}, L109 (1999).

\bibitem{FAST:2019zow}
J.~Lu \textit{et al.} [FAST],
Sci. China Phys. Mech. Astron. \textbf{62}, no.5, 959505 (2019)

\bibitem{Lai:2017xys}
X.~Y.~Lai, C.~A.~Yun, J.~G.~Lu, G.~L.~L\"u, Z.~J.~Wang and R.~X.~Xu,
Mon. Not. Roy. Astron. Soc. \textbf{476}, no.3, 3303-3309 (2018).

\bibitem{Wang:2020xsm}
W.~Wang, X.~Lai, E.~Zhou, J.~Lu, X.~Zheng and R.~Xu,
Mon. Not. Roy. Astron. Soc. \textbf{500}, no.4, 5336-5349 (2020).
%
\bibitem{Zhou:2004ue}
A.~Z.~Zhou, R.~X.~Xu, X.~J.~Wu, N.~Wang and X.~Y.~Hong,
Astropart. Phys. \textbf{22}, 73-79 (2004).

\bibitem{Peng:2007eq}
C.~Peng and R.~X.~Xu,
Mon. Not. Roy. Astron. Soc. \textbf{384}, 1034-1038 (2008).

\bibitem{Zhou:2014tba}
E.~P.~Zhou, J.~G.~Lu, H.~Tong and R.~X.~Xu,
Mon. Not. Roy. Astron. Soc. \textbf{443}, no.3, 2705-2710 (2014).


\bibitem{Gao:2021uus}
Y.~Gao, X.~Y.~Lai, L.~Shao and R.~X.~Xu,
Mon. Not. Roy. Astron. Soc. \textbf{509}, no.2, 2758-2779 (2021).

\bibitem{Li:2022qql}
H.~B.~Li, Y.~Gao, L.~Shao, R.~X.~Xu and R.~Xu,
Mon. Not. Roy. Astron. Soc. \textbf{516}, 6172 (2022).

\bibitem{Ouyed:2023hqe}
R.~Ouyed, D.~Leahy, N.~Koning and P.~Jaikumar,
[arXiv:2302.06820 [astro-ph.CO]].

%
%
%
\bibitem{Carr1974} B. J. Carr and S. W. Hawking, 
Mon. Not. Roy. Astr. Soc. \textbf{168}, 399 (1974).
\bibitem{Nadezhin1978} D. K. Nadezhin, I. D. Novikov and A. G. Polnarev, 
Astron. Zh. \textbf{55}, 216 (1978).
%
\bibitem{Farrar2003} G. R. Farrar,
Int. J. Theor. Phys. \textbf{42}, 1211-1218 (2003).
%
\bibitem{Farrar2022}
G. R. Farrar, 
arXiv:2201.01334.
%
\bibitem{Flambaum2021}
V. V. Flambaum and I. B. Samsonov, 
arXiv:2112.07201.
%
\bibitem{Alonso-Alvarez2021}
G. Alonso-\~Alvarez, G. Elor, M. Escudero, B. Fornal, B. Grinstein, and J. M. Camalich, 
arXiv:2111.12712.
%
\bibitem{Zhitnitsky2003}
A. R. Zhitnitsky, 
J. Cosmol. Astropart. P. \textbf{10}, 010 (2003).
%
\bibitem{Guenther:2020jwe}
J.~N.~Guenther,
Eur. Phys. J. A \textbf{57}, no.4, 136 (2021).
%
\bibitem{Bzdak:2019pkr}
A.~Bzdak, S.~Esumi, V.~Koch, J.~Liao, M.~Stephanov and N.~Xu,
Phys. Rept. \textbf{853}, 1-87 (2020).
%
\bibitem{Fischer:2018sdj}
C.~S.~Fischer,
Prog. Part. Nucl. Phys. \textbf{105}, 1-60 (2019).
%
\bibitem{Lineweaver2008} C. H. Lineweaver, C. Egan,
Phys. Life Rev. \textbf{5}, 225-242 (2008).
%
\bibitem{Zemansky1997}  M. W. Zemansky, R. H. Dittman,
Heat and Thermodynamics, 1st ed.; 
McGraw-Hill: New York, NY, USA, (1997).
%
\bibitem{Planck} Planck Collaboration, Aghanim, N., Akrami, Y., et al.\ 2020, Astron. \& Astrophys., 641, A6. 
%
\bibitem{Rehberg1996} P. Rehberg, S. P. Klevansky, and J. H\"ufner,
Phys. Rev. C \textbf{53}, 410 (1996).
%
\bibitem{Ratti2006} C. Ratti, M. A. Thaler, and W. Weise, 
Phys. Rev. D \textbf{73}  014019 (2006).
%
\bibitem{Glendenning1991} N. K. Glendenning and S. A. Moszkowski,
Phys. Rev. Lett. \textbf{67} 2414-2417 (1991). 
%
\bibitem{Lai2017}  X. Y. Lai and R. X. Xu, 
Journal of Physics Conf. Series, \textbf{861}, 012027 (2017).
%
\bibitem{Wang2018} Z. Wang,  J. G. Lu, and R. X. Xu,  
JPS Conf. Proc., \textbf{20}, 011032 (2018).
%
\bibitem{Masuda2013a} K. Masuda, T. Hatsuda, and T. Takatsuka,
Prog. Theor. Exp. Phys. 073D01 (2013).
\bibitem{Masuda2013b} K. Masuda, T. Hatsuda, and T. Takatsuka,
Astrophys. J \textbf{764}, 12 (2013).
\bibitem{Masuda2016a} K.Masuda, T.Hatsuda, and T.Takatsuka, 
Prog. Theor. Exp. Phys.  021D01 (2016).
\bibitem{Masuda2016b} K.Masuda, T.Hatsuda, and T.Takatsuka, 
Eur. Phys. J. A \textbf{52}, 65 (2016).
\bibitem{Whittenbury2016} D. L. Whittenbury, H. H. Matevosyan, and A. W. Thomas, 
Phys. Rev. C \textbf{93}, 035807 (2016).
%
\bibitem{Li2018} C.-M. Li, Y. Yan, J.-J. Geng, Y.-F. Huang, H.-S. Zong, 
Phys. Rev. D \textbf{98}, 083013 (2018).
%
\bibitem{Kojo2016} T. Kojo, P. D. Powell, Y. Song, and G. Baym, 
Nucl. Phys. A \textbf{956}, 821 (2016).



\end{thebibliography}
\end{document}